\documentclass{svjour3}

\usepackage{microtype}
\usepackage{mathptmx}

\usepackage{changes}
\usepackage{graphicx}
\graphicspath{{figures/}}
\usepackage{natbib}
\bibpunct{(}{)}{;}{a}{}{,}

\usepackage{Macros}

\begin{document}

\title{Intensification of tilted atmospheric vortices by asymmetric diabatic heating}

\author{Tom Dörffel \and %
        Ariane Papke \and %
        Rupert Klein \and %
        Natalia Ernst \and %
        Piotr K.\ Smolarkiewicz}

\institute{T. Dörffel and A. Papke and R. Klein \at
              FB für Mathematik und Informatik, Freie Universität Berlin, Berlin, Germany\\
              \email{rupert.klein@fu-berlin.de}
          \and
          N. Ernst \at
              Zuse Insitute Berlin, Berlin, Germany \\
          \and
          P. Smolarkiewicz \at
              National Center for Atmospheric Research, Boulder, CO 80307, USA
}

\date{Received: date / Accepted: date}

\maketitle

%
\begin{abstract}%
\citet{PaeschkeEtAl2012} studied the nonlinear dynamics
of strongly tilted vortices subject to asymmetric diabatic heating by asymptotic methods.
They found, \ia, that an azimuthal Fourier mode~1 heating pattern can intensify or
attenuate such a vortex depending on the relative orientation of tilt and heating
asymmetries. The theory originally addressed the gradient wind regime which,
asymptotically speaking, corresponds to vortex Rossby numbers of order unity
in the limit. Formally, this restricts the applicability of the theory to rather weak
vortices in the near equatorial region. It is shown below that said theory is, in
contrast, uniformly valid for vanishing Coriolis parameter and thus applicable to
vortices up to low hurricane strengths. In addition, the paper presents an
extended discussion of the asymptotics as regards their physical interpretation and
their implications for the overall vortex dynamics. The paper's second
contribution is a series of three-dimensional numerical simulations examining the effect
of different orientations of dipolar heat release on idealized tropical cyclones.
Comparisons with numerical solutions of the asymptotic equations yield evidence that supports the original predictions. In addition, the influence of asymmetric diabatic heat release on the time evolution of centerline tilt is analysed further, and a steering mechanism based on the orientation of the heating dipole is revealed.
\end{abstract}


\section{Introduction}
\label{sec:Intro}

Atmospheric vortex intensification and the associated evolution of vortex structure remain
a topic of intense investigations. As \citet{SmithMontgomery2017} point out in their
review article, intricate interactions of boundary layer processes, moist thermodynamics,
multiscale stochastic deep convection, and the vortex-scale fluid dynamics produce the
observed, sometimes extremely rapid intensification of incipient hurricanes. They also
emphasize that, despite the valuable insights that have been gained in many studies of
idealized axisymmetric flow models, asymmetries of vortex structure, convection patterns,
and boundary layer structure have been observed to be important for vortex intensification
in real-life situations.

This study focuses on the principal response mechanisms of \emph{Tropical cyclone}-like (TC) atmospheric vortices
to asymmetric diabatic heat release following the theory of \citet{PaeschkeEtAl2012}.
Therefore, we analyze both the structure and intensity of the bulk vortex above the
boundary layer under the influence of different configurations of asymmetric heating.
From \citet{NolanMontgomery2002,NolanGrasso2003,NolanEtAl2007}, among others, we adopt
the point of view that latent heat release from condensation can be modeled, with
limitations, by external diabatic heat sources in dry air. In the cited studies,
non-axisymmetric heating patterns were shown to have at most a small effect on vortex
strength within the framework of linearizations about an axisymmetric upright vortex.
These results of linear theory were corroborated in \citet{NolanGrasso2003,NolanEtAl2007}
by comparison with fully nonlinear three-dimensional simulations.
By both, analytical and numerical examination, we will see that the particular flow
structure of a strongly tilted vortex allows for a leading order intensification mechanism
based on asymmetric heating that cannot be captured for linearized (weak) vortex tilts.

Investigating incipient hurricanes that develop in the tropical
Atlantic, \citet{MarksHouzeGamache1992,Marks2003,DunkertonEtAl2009} revealed,
that such vortices can exhibit
very strong tilt. Thus, for instance, the locations of the vortex center at heights
equivalent to the $\unit{925}{\hecto\pascal}$ and $\unit{200}{\hecto\pascal}$ pressure
levels are located about $\unit{200}{\kilo\meter}$ apart, e.g., in
\citep[][figure~18]{Marks2003} and \citep[][figure~21]{DunkertonEtAl2009}.
This amounts to an overall vortex tilt at a scale comparable to the vortex diameter,
\ie, to a situation that clearly does not allow for linearizations about an upright
vortex. In fact, in characterizing the wind field of Hurricane Norbert,
\citet{MarksHouzeGamache1992} already utilized the concept of a height-dependent
vortex center, i.e., of time dependent centerline, and this is one of the key
structural aspects in the analysis of \citet{PaeschkeEtAl2012}, which we revisit in
this paper.

\citet{PaeschkeEtAl2012} analyzed the
dynamics of such strongly tilted atmospheric vortices in the gradient wind regime by
matched asymptotic expansions. They obtained a closed coupled set of evolution
equations for the primary circulation structure and the vortex centerline, and
demonstrated that in a strongly tilted vortex symmetric and asymmetric heating
patterns can have a comparable impact on vortex intensity. As by its very definition
the gradient wind regime is restricted to vortex Rossby numbers of order unity,
this theory has thus far been considered applicable only to rather weak vortices
with intensities relatively far from the interesting stage of the tropical
storm/hurricane transition \citep{MontgomeryPrivate2017}.

\begin{figure}
\begin{center}
\includegraphics{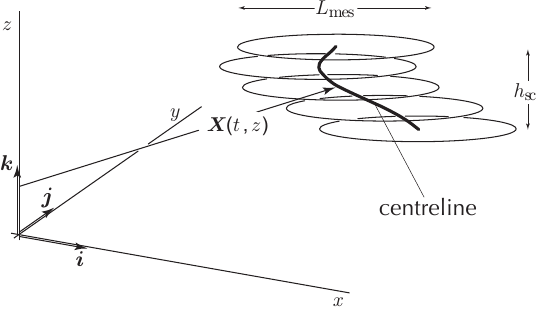}
\caption{Sketch of the spacial scaling regime for vortices in this work.
In each horizontal plane the vortex is axisymmetric to leading order while
the vortex center line position $\XC(t,z)$ covers horizontal distances
comparable to the vortex core size (adapted from \citet{PaeschkeEtAl2012}).}
\label{fig:TiltedVortex}
\end{center}
\end{figure}

To allow for vortices in this transition regime, we consider here in the
first part of the paper the dynamics
of meso-scale atmospheric vortices $\Lmes \sim \unit{100}{\kilo\meter}$ that extend
vertically across the depth of the troposphere $\hsc \sim \unit{10}{\kilo\meter}$
but feature large vortex Rossby number $\Romes \gg 1$. We use the asymptotic
techniques introduced by \citet{PaeschkeEtAl2012} and recycle many of their technical
steps. As indicated in fig.~\ref{fig:TiltedVortex}, we assume vortices with nearly
axisymmetric core structure at each horizontal level, and we allow for strong vortex
tilt such that the vortex centers observed at different heights may be displaced
horizontally relative to each other by distances comparable to the vortex core
size $\Lmes$.

One of the main findings of \citet{PaeschkeEtAl2012} was the following evolution
equation for the primary circulation described by the axisymmetric leading-order
circumferential velocity, $\uthet$, valid for time scales large compared
to the vortex turnover time scale,
\begin{equation} \label{eq:TempevoluthetaIntro}
\pderiv{u_{\theta}}{t}
+ w_{0} \pderiv{u_{\theta}}{z}
+ u_{r,00}
  \left(\pderiv{u_{\theta}}{r} + \frac{u_{\theta}}{r} + f_0
  \right)
= - u_{r,*}
  \left(\frac{u_{\theta}}{r} + f_0
  \right)\, .
\end{equation}
Here $(t,r,z)$ are the appropriately rescaled time, radial, and vertical coordinates,
$f_0$ is the Coriolis parameter, and $w_0$ and $u_{r,00}$ are the axisymmetric components of
the vertical and radial velocities induced by \emph{the axisymmetric mean heating patterns},
also properly rescaled. The apparent radial velocity
$u_{r,*}$ results from an interaction of the vortex tilt with the asymmetric first
circumferential Fourier mode of the vertical velocity. In particular,
\begin{equation}\label{eq:UrStar_intro}
u_{r,*}
= \frac{1}{2\pi}\int\limits_{-\pi}^{\pi} w \, \verad \cdot\pderiv{\XC}{z} \, d\theta
 \, ,
\end{equation}
where $\XC(t,z)$ is the time dependent vortex centerline position at height $z$
(see fig.~\ref{fig:TiltedVortex}), $w$~is the full vertical velocity, and
$\verad = \vi \cos(\theta) + \vj \sin(\theta)$ is the radial unit vector of a
horizontal polar coordinate system attached to the centerline.

$\vect X(t,z)$ itself is governed by the \emph{centerline equation}:
\begin{equation}
 \pderiv{\vect X}{t} = \vect u_s + (\vect X \cdot \nabla) \vect u_s + \ln\frac{1}{\delta} \vect k \times \vect M_1 + \vect k \times \vPsi
 \label{eq:TempevolutionCenterlineIntro}
\end{equation}
$\vect u_s(t,z)$ expresses the background wind profile, $\delta$ is a small number according to the asymptotic scaling, $\vect k$ the vertical unit vector, $\vect M_1$ is a weighted curvature measure of $\vect X$ and $\vPsi$ evaluates Fourier-1 modes of vertical velocity resulting from both, diabatic heating and adiabatic balances within a tilted vortex.
In the adiabatic case and without vertical wind shear equation \eqref{eq:TempevolutionCenterlineIntro} simplifies to a linear Schrödinger-like equation exhibiting undamped precession of eigenmodes.
More details on the expressions for $\vect M_1$ and $\vPsi$ follow in the further course of this article.

The main findings of present work are:
\begin{enumerate}

\item \label{item:IntroA}
The evolution equation from \eq{eq:TempevoluthetaIntro} is uniformly valid
as $f_0 \to 0$ so that it holds, in particular, also for $\Romes \gg 1$, \ie, for
vortices of hurricane strength.
In fact, we argue the for $\Romes\geq1$ the structure of the leading-order equations does not change.

\item \label{item:IntroB}
The mechanism of vortex spin-up by asymmetric heating of a tilted vortex is
traced back analytically to an effective circumferential mean vertical mass
flux divergence that arises when the first Fourier mode diabatic heating
and the vortex tilt correlate negatively.

\item Asymmetric heating introduces a forcing of the vortex motion which can accelerate/decelerate the centerline precession and/or increase/decrease its tilt depending on the relative orientation of tilt and heating dipole.

\item \label{item:IntroC}
Equation \eq{eq:TempevoluthetaIntro} can be recast into a balance equation
for kinetic energy, $\ekin = \rho_0\frac{\uthet^2}{2}$,
\begin{equation}\label{eq:KineticEnergyBudget}
\dss \left(r \ekin\rule{0pt}{9pt}\right)_t
+ \left(r u_{r,00} \left[\ekin + \widetilde p\right]\rule{0pt}{9pt}\right)_{r}
+ \left(r \wc_0 \left[\ekin + \widetilde p\right]\rule{0pt}{9pt}\right)_z
= \dss \ \frac{r\rho_0}{N^2\Thetabar^2}
     \left(\widetilde\Theta \cdot Q_{\Theta} \right)_0
\end{equation}
in line with the theory by \citet{Lorenz1955} for available potential energy (APE) generation.

Here $\widetilde p$ is the relevant pressure perturbation,
$\widetilde{\Theta}, Q_{\Theta}$ are the potential temperature perturbations and the diabatic heating, respectively, and $(\cdot)_0$ corresponds to the axisymmetric mean.
In the current case, it encodes the correlation of potential temperature perturbation and diabatic heating.
$N$ and $\Thetabar$ are the Brunt-V\"ais\"al\"a frequency
and the background potential temperature stratification, respectively.
Equation \eq{eq:KineticEnergyBudget} states that, except for a conservative
redistribution of kinetic energy due to advection and the work of the pressure
perturbation, $\tilde p$, positive correlations of diabatic sources and potential temperature
perturbations generate the potential energy available for increasing the kinetic
energy of the vortex.

\citet{NolanEtAl2007} study the effects of asymmetric diabatic heating on vortex
strength in a linearized model. One of their conclusions is that ``... purely
asymmetric heating generally leads to vortex weakening, usually in terms of the
symmetric energy, and always in terms of the low-level wind.'' The present theory
shows that this conclusion does not hold up in case of a strongly tilted vortex,
but that in this case symmetric and suitably arranged asymmetric heating have
vortex intensification efficiencies of the same order of magnitude.

\item \label{item:IntroD}
The theory compares favorably with three-dimensional numerical simulations based on
the compressible Euler equations.

\end{enumerate}

To arrive at these results, we first recount the governing equations and the principles
of our analytical approach in section~\ref{sec:Methodology}, and then revisit the
derivations by \citet{PaeschkeEtAl2012}.
A discussion of the scaling regime is given in section~\ref{sec:MesoScaleAnalysis} to
investigate the influence of the Coriolis effect (item~(\ref{item:IntroA})), and the asymptotic vortex core expansion is carried out in section \ref{sec:CoreStructureEvolution} analytically supporting the physical interpretation of the asymmetric intensification mechanism given in item~(\ref{item:IntroB}).
In section \ref{sec:Discussion} we establish the kinetic energy balance of item~(\ref{item:IntroC}).
Section~\ref{sec:Simulations} presents results of the theory in comparison with three-dimensional computational simulations to corroborate item~(\ref{item:IntroD}). Conclusions and an outlook are provided in
section~\ref{sec:Conclusions}.


\section{Dimensionless governing equations and distinguished limits}
\label{sec:Methodology}


\subsection{Governing equations}
\label{ssec:GoverningEquations}

The dimensionless inviscid rotating compressible flow equations for an ideal gas with
constant specific heat capacities in the beta plane approximation form the basis
for the subsequent asymptotic analysis:
\begin{subequations}
\label{eq:CompressibleFlowEquationsDimless}
\begin{alignat}{7}
&\pp{\vp}{t}
  &
    & + \vp\cdot\gradhor{\vp}
      &
        & + w \, \pp{\vp}{z}
          &
            & + \frac{1}{\M^2} \frac{1}{\rho} \, \gradhor{p}
              &
                & + \frac{1}{\Ro}\left(1 + \widehat\beta y\right)\, \vect{k} \times {\vp}
                  &
                    & = 0\, ,
                      \\
&\pp{w}{t}
  &
    & + \vp \cdot \gradhor{}\statt{\vp}{w}
      &
        & + w\, \pp{w}{z}
          &
            & + \frac{1}{\M^2}\frac{1}{\rho}\, \pp{p}{z}
              &
                &
                  &
                    & = -\frac{1}{\Fr^2} ,
                      \\
\label{eq:ContinuityOriginal}
&\pp{\rho}{t}
  &
    & + \vp\cdot\gradhor{\rho}
      &
        & + w \, \pp{\rho}{z}
          &
            & + \rho \gradhor\cdot \vp
              &
                & + \rho \pp{w}{z}
                  &
                    & = 0\, ,
                      \\
&\pp{\Theta}{t}
  &
    & + \vp\cdot\gradhor{\Theta}
      &
        & + w\, \pp{\Theta}{z}
          &
            &
              &
                &
                  &
                    & = Q_{\Theta} \, ,
                      \\\label{eq:PotTempDimensional}
&
  &
    &
      &
        &
          &
            &
              &
                &
                  & \rho\Theta
                    & = p^{\frac{1}{\gamma}}
\end{alignat}
\end{subequations}
Here $p,\rho,\Theta,\vp,w$ are rescaled pressure, density, potential temperature, and the
horizontal and vertical velocities, and $\gamma$ is the specific heat ratio.

The three-dimensional
gradient is $\grad = \gradhor{} + \vk\,\partial/\partial z$ with the horizontal gradient
$\gradhor{} = \vi\, \partial/\partial x + \vj\, \partial/\partial y$, the
zonal, meridional, and vertical coordinates $(x,y,z)$, and the related unit vectors
$(\vi,\vj,\vk)$. Finally, $t$ is the time variable and $Q_{\Theta}$ is a
diabatic source term.

Table \ref{tab:AtmosphericCharacteristics} lists general characteristics of the near-tropical atmosphere.
Together with the combined values in Table~\ref{tab:ReferenceValues} they form reference values for non-dimensionalization.
Let an asterisk denote dimensional quantities, then the unknowns and coordinates in \eq{eq:CompressibleFlowEquationsDimless} are
\begin{equation}\label{eq:DimensionlessVariables}
p = \frac{p^*}{\rfr{p}}\, ,
\quad
\rho = \frac{\rho^*}{\rfr{\rho}}\, ,
\quad
(\vp, w) = \frac{(\vp^* , w^*)}{\rfr{u}}\, ,
\quad
(\vx, z) = \frac{(\vx^*,z^*)}{\hsc} \, ,
\quad
t = \frac{t^*\rfr{u}}{\hsc}\,.
\end{equation}
Note that $\rfr{u}/\hsc$ is an estimate of the large-scale thermal wind shear, and $\vx = \vi x + \vj y$ is the horizontal coordinate vector.
\begin{table}
  \begin{center}
    \begin{tabular}{@{}llcrl@{\ \ }rl}
      Gravitational acceleration
        & $g$
          & $=$
            &
              &
                & $9.81$
                  & $\meter\,\second^{-2}$
                    \\[0pt]
      Coriolis parameter ($\phi = 30\degree$ N)
        & $\rfr{f}$
          & $=$
            &
              &
                & $7.3\cdot 10^{-5}$
                  & $\second^{-1}$
                    \\[0pt]
       $(df/dy)_0$ ($\phi = 30\degree$ N)
        & $\rfr{\beta}$
          & $=$
            &
              &
                & $2.0\cdot 10^{-11}$
                  & $\meter^{-1}\,\second^{-1}$
                    \\[0pt]
      Pressure
        & $\rfr{p}$
          & $=$
            &
              &
                & $10^5$
                  & $\pascal$
                    \\[0pt]
      Temperature
        & $\rfr{T}$
          & $=$
            &
              &
                & $300$
                  & $\kelvin$
                    \\[0pt]
      Brunt-V\"ais\"al\"a frequency
        & $\rfr{N}$
          & $=$
            &
              &
                & $10^{-2}$
                  & $\second^{-1}$
                    \\[0pt]
      Dry air gas constant
        & $R$
          & $=$
            &
              &
                & $287$
                  & $\meter^2\,\rpsquare\second\,\kelvin^{-1}$
                    \\[0pt]
      Isentropic exponent
        & $\gamma$
          & $=$
            &
              &
                & $1.4$
                  &
                    \\[0pt]
      \end{tabular}
    \end{center}
    \caption{Characteristic atmospheric flow parameters.
    \label{tab:AtmosphericCharacteristics}}
\end{table}
\begin{table}
  \begin{center}
    \begin{tabular}{@{}l@{\qquad}lcrl@{\ \ }rl}
      Density
        & $\rfr{\rho}$
          & $=$
            & $\dss \frac{\rfr{p}}{R\rfr{T}}$
              & $\sim$
                & $1.16$
                  & \kilo\gram\,\meter$^{-3}$
                    \\[5pt]
      Potential temperature
        & $\Delta \Theta$
          & $=$
            & $\dss \rfr{T}\frac{\hsc \rfr{N}^2}{g}$
              & $\sim$
                & $40$
                  & $\kelvin$
                    \\[5pt]
      Velocity
        & $\rfr{u}$
          & $=$
            & $\dss \frac{\tan\phi}{\pi/2}\frac{\rfr{N}^2}{\rfr{f}^2}\beta\hsc^2$
              & $\sim$
                & $10$
                  & \meter\,\second$^{-1}$
                    \\[5pt]
      Length
        & $\hsc$
          & $=$
            & $\dss \frac{\rfr{p}}{g\rfr{\rho}}$
              & $\sim$
                & $8.8$
                  & \kilo\meter
                    \\[5pt]
      Time
        & $\rfr{t}$
          & $=$
            & $\dss \frac{\hsc}{\rfr{u}}$
              & $\sim$
                & $10^3$
                  & \second \\
      \end{tabular}
    \end{center}
    \caption{Derived reference values for non-dimensionalization.
    \label{tab:ReferenceValues}}
\end{table}

In deriving the dimensionless equations \eq{eq:CompressibleFlowEquationsDimless}
using the quantities from tables~\ref{tab:AtmosphericCharacteristics} and
\ref{tab:ReferenceValues} the Mach, internal wave Froude, and Rossby numbers, and the $\beta$-parameter
\begin{equation}
\label{eq:GeophysicalParameters}
\begin{array}{lcccl}
\M
  & =
    & \dss \frac{\rfr{u}}{\sqrt{ R \rfr{T}}}
      & \approx
        & 3.4\cdot 10^{-2}
          \\
\Fr
  & =
    & \dss \frac{\rfr{u}}{\rfr{N}\hsc}
      & \approx
        & 1.1\cdot 10^{-1}
\end{array}\,,
\qquad
\begin{array}{lcccl}
\Ro
  & =
    & \dss \frac{\rfr{u}}{\rfr{f} \hsc}
      & \approx
        & 13.3
          \\
\betahat
  & =
    & \dss \frac{\rfr{\beta} \hsc}{\rfr{f}}
      & \approx
        & 2.7\cdot10^{-3}
\end{array}
\end{equation}
emerge naturally.
These are replaced with functions of a single small expansion parameter
$\eps \ll 1$ through the distinguished limits
\begin{equation}\label{eq:distinguished_limits}
\M = \eps^{3/2}\, ,
\quad
\Fr = \frac{\eps}{N}\, ,
\quad
\Ro = \frac{1}{\eps f}\, ,
\quad
\betahat = \eps^3 \beta\, ,
\end{equation}
in line with the multiscale asymptotic modelling framework of \citet{Klein2010}.
Here $(N, f, \beta) = \bigo{1}$ as $\eps\to 0$, with
concrete values
\begin{equation}
N = 0.91\,,
\quad
f = 0.75\,,
\quad
\beta = 2.7
\end{equation}
derived from \eq{eq:GeophysicalParameters} for $\eps = \M^{2/3} = 0.105$.
Replacing the characteristic numbers according to \eqref{eq:distinguished_limits} we get
\begin{subequations}
\label{eq:AsymptoticFlowEquations}
\begin{alignat}{7}
&\pp{\vp}{t}
  &
    & + \vp\cdot\gradhor{\vp}
      &
        & + w \, \pp{\vp}{z}
          &
            & + \frac{1}{\eps^3} \frac{1}{\rho} \, \gradhor{p}
              &
                & + \eps \left(f + \eps^3\beta y\right)\, \vect{k} \times {\vp}
                  &
                    & = 0\, ,
                      \\
&\pp{w}{t}
  &
    & + \vp \cdot \gradhor{}\statt{\vp}{w}
      &
        & + w\, \pp{w}{z}
          &
            & + \frac{1}{\eps^3}\frac{1}{\rho}\, \pp{p}{z}
              &
                &
                  &
                    & = - \frac{1}{\eps^3} ,
                      \\
\label{eq:ContinuityDistLim}
&\pp{\rho}{t}
  &
    & + \vp\cdot\gradhor{\rho}
      &
        & + w \, \pp{\rho}{z}
          &
            & + \rho \gradhor\cdot \vp
              &
                & + \rho \pp{w}{z}
                  &
                    & = 0\, ,
                      \\
&\pp{\Theta}{t}
  &
    & + \vp\cdot\gradhor{\Theta}
      &
        & + w\, \pp{\Theta}{z}
          &
            &
              &
                &
                  &
                    & = Q_{\Theta} \, ,
                      \\\label{eq:PotTempDistLim}
&
  &
    &
      &
        &
          &
            &
              &
                &
                  & \rho\Theta
                    & = p^{\frac{1}{\gamma}} \, .
\end{alignat}
\end{subequations}
Whereas
$f$ and $\beta$ appear explicitly in \eqref{eq:AsymptoticFlowEquations}, $N$ characterizes the background stratification of potential temperature and will be invoked below where we define the initial conditions for the vortex flow.

Equations \eq{eq:AsymptoticFlowEquations} will form the basis for the
subsequent asymptotic analysis for $\eps \ll 1$, although much of the expansions
will proceed in terms of the small parameter
\begin{equation}\label{eq:DeltaDefinition}
\sqeps = \sqrt{\eps}\,.
\end{equation}
%


\section{Scaling regime for large vortex Rossby number and strong tilt}
\label{sec:MesoScaleAnalysis}


\subsection{Vortex core size, intensity, and evolution time scale}
\label{ssec:CoreScalings}

Vortex core sizes of $\unit{50}{\kilo\meter}$ to $\unit{200}{\kilo\meter}$ are typical
for tropical storms and hurricanes, and the storm/hurricane threshold lies at wind
speeds of $\unit{30}{\meter\per\second}$ \citep{Emanuel2003}.
With $\epsonehalf^2 \equiv \eps \sim 1/10$, $\hsc \sim \unit{10}{\kilo\meter}$, and
$\rfr{u} \sim \unit{10}{\meter\per\second}$, these data correspond well with
\begin{equation}\label{eq:CoreScalings}
\Lvort \sim \hsc/\epsonehalf^2 \approx \unit{100}{\kilo\meter}\,,
\quad
\umax \sim  \rfr{u}/\epsonehalf \approx \unit{33}{\meter\per\second}\, ,
\quad
\epsonehalf p_{\rm v} \sim \epsonehalf^4\rfr{p}\,,
\end{equation}
%
for a characteristic vortex core size $\Lvort$, a typical wind speed, and the associated
depression in the vortex core, respectively. Note that these scalings deviate from
those adopted by \citet{PaeschkeEtAl2012}, who considered systematically larger
radii of the order $\Lvort \sim \hsc/\epsonehalf^{3}$ needed for direct matching to a
quasi-geostrophic large scale outer flow. From their work we recall, however, that the
vortex core structure and tilt develop on a time scale $t_{\rm v}$ that is by
$1/\epsonehalf^2$ longer than the vortex core turnover time scale
$t_{\rm to} = \Lvort / \umax$. Thus, in view of \eq{eq:CoreScalings}, we will follow
the vortex core evolution on the time scale
\begin{equation}\label{eq:TimeScales}
t_{\rm v}
= \frac{t_{\rm to}}{\epsonehalf^2}
= \frac{1}{\epsonehalf^2}\frac{\hsc}{\epsonehalf^2}\frac{\epsonehalf}{\rfr{u}}
= \frac{\rfr{t}}{\epsonehalf^{3}} \sim \unit{10}{\hour}\,.
\end{equation}
The scalings in \eq{eq:CoreScalings} and \eq{eq:TimeScales} include the regime of
``rapid intensification'', defined by NOAA's National Hurricane Center%
\footnote{http://www.nhc.noaa.gov/aboutgloss.shtml} to denote maximum wind accelerations
of $30\, {\rm kt} \sim \unit{15}{\meter\per\second}$ in $24 \hour$.

Also, the adopted scalings describe a vortex in the cyclostrophic
regime since
\begin{equation}\label{eq:CyclosStrophicRegime}
\frac{\hsc}{\rfr{u}^2} \frac{\uthet^2}{r} = \bigoh{1}
\qquad\text{whereas}\qquad
\frac{\hsc}{\rfr{u_\theta}^2} \rfr{f} \uthet = \frac{1}{\Ro}\frac{u}{\rfr{u}} = \bigoh{\epsonehalf} \,,
\end{equation}
\ie, the Coriolis term is subordinate to the centripetal acceleration in the horizontal
momentum balance in this regime. Accordingly, the vortex Rossby number is large,
\begin{equation}
\Ro_{\rm v}
= \frac{u_{\rm max}}{f_0\Lvort}
= \Ro \frac{u_{\rm max}}{\rfr{u}} \frac{\hsc}{\Lvort}
= \bigoh{\epsonehalf^{-2-1+2}}
= \bigoh{\frac{1}{\epsonehalf}}\,.
\end{equation}
%


\subsection{Co-moving coordinates for a strongly tilted vortex}
\label{ssec:TiltCoordinates}

Following \citet{PaeschkeEtAl2012}, we resolve the flow dynamics on the vortex
precession and core evolution time scale $t_{\rm v}$ from \eq{eq:TimeScales}. The
appropriate time coordinate is
\begin{equation}
\tv = \epsonehalf^3 t\, .
\end{equation}
For the core structure analysis we introduce vortex centered horizontal coordinates
\begin{equation}
\vx
=
\frac{1}{\sqeps^2}\bigl(\XC(\tv, z) + \vxhat\bigr)
\end{equation}
where $\XC(\tv, z)$ is the horizontal position of the vortex centerline at height $z$
and $\vxhat$ is the relative horizontal offset. With this scaling $\vxhat$ resolves
the core scale $\Lvort$ from \eq{eq:CoreScalings} and the centerline covers
comparable distances. This justifies the notion of ``strong tilt''.

In the sequel we use polar coordinates in horizontal planes, \ie,
\begin{equation}\label{eq:AttachedCoordinates_Ia}
\vxhat = \widehat{x} \ \vi\ + \widehat{y} \ \vj
\qquad\hbox{where}\qquad
\left\{
\begin{array}{l}
\dss \widehat{x} = \rmeso \cos \theta \, ;
      \\
\dss\widehat{y} =  \rmeso \sin \theta\, ;
\end{array}
\begin{array}{l}
\dss\vi = \verad \cos \theta - \vetheta \sin \theta
      \\
\dss\vj = \verad \sin \theta + \vetheta \cos \theta
\end{array}
\right.
\end{equation}
with $\verad $ and $\vetheta$ the radial and circumferential unit vectors, respectively.
The transformation rules for derivatives in these coordinates read
\begin{subequations}\label{eq:DerivativeTransformations}
\begin{align} \label{horderiv}
  \gradhor \ \
    & = \ \ \sqeps^{2} \left(\verad \pderiv{}{\rmeso}
          + \vetheta \frac{1}{\rmeso} \pderiv{}{\theta} \right)  \ \
      \equiv \ \  \sqeps^{2}\, \nablac\ ,
      \\[5pt] \label{vertderiv}
  \pderiv{}{z}\bigg \vert_{t,x,y}
    & = \ \ \pderiv{}{z} \bigg \vert_{\tv,\rmeso,\theta}
           - \ \tilt\cdot\nablac
           \ ,
      \\[5pt]  \label{timederiv}
  \pderiv{}{t}\bigg \vert_{x,y,z}
    &= \ \delta^3 \left( \pderiv{}{\tv}\bigg \vert_{\rmeso,\theta,z} \
        - \  \pderiv{\XC}{\tv}\cdot \nablac \right)\, .
\end{align}
\end{subequations}
The horizontal velocity is decomposed into the vortex' motion plus the relative velocity,
\begin{equation}
  \label{HorizontalVelocitySplit}
  \vp = \sqeps \pderiv{\XC}{\tv} + (u_r\, \verad + u_{\theta}\, \vetheta)\, .
\end{equation}
For later reference, here are the centerline represented in the $(\verad, \vetheta)$ basis,
\begin{equation} \label{eq:CenterlinePolar}
\XC
=
\left(X\cos \theta   + Y\sin \theta\right) \ \verad
   + \left(- X \sin \theta + Y \cos \theta\right) \ \vetheta \, ,
\end{equation}
and the Fourier expansion of functions of the circumferential angle,
$\theta$,
\begin{equation}\label{eq:Fourier}
F(\theta) = F_0 + \sum_n \left(F_{n1} \cos(n\theta) + F_{n2} \sin(n\theta)\right) \,.
\end{equation}
Note that we have exchanged the roles of $F_{n1}$ and $F_{n2}$ relative to their use
in \citep{PaeschkeEtAl2012} as this will streamline the analysis of the orientation of a dipolar field characterized by $\vect F_1 = (F_{11}, F_{12})^T$ relative to the tilt vector $\partial \vect X /\partial z$.


\subsection{Vortex core expansion scheme}
\label{ssec:SpecExpansionmeso}

The circumferential velocity is expanded as
\begin{subequations}\label{Meso:HorizontalVelocity}
\begin{alignat}{6}
  \label{Meso:HorizontalVelocity1}
  &\uthet(t,\vx,z;\varepsilon) \ \
  && = \ \ \epsminusonehalf\uthet\order{0}(t, \rmeso, z)
  && + \uthet\order{1}(t, \rmeso, z)
  && + \epsonehalf\uthet\order{2}(t, \rmeso,\theta, z)
  && + \littleo{\epsonehalf}\, ,
  \\ \label{Meso:HorizontalVelocity2}
  &\urad(t,\vx,z;\varepsilon) \ \
  && = \ \
  &&
  && \phantom{+ .}\: \sqeps\, \urad\order{2}(t, \rmeso,\theta, z)
  && + \littleo{\epsonehalf}\, .
\end{alignat}
\end{subequations}
\ie, non-axisymmetry relative to the centerline is allowed for scaling orders from
$\bigo{\epsonehalf\rfr{u}}$ upwards. Across the core size length scale, $\Lvort$,
such asymmetries induce horizontal divergences of order
$\urad / \Lvort \sim \epsonehalf \rfr{u} / (\hsc / \epsonehalf^2) = \sqeps^3 \rfr{u}/\hsc$,
see~\eq{eq:CoreScalings}. Since the flow field is anelastic to leading order
as derived below, this implies the vertical velocity scaling,
\begin{alignat}{6}
  \label{Meso:VerticalVelocity}
  w(t,\vx,z;\varepsilon)
  = \epsonehalf^3 \wc\order{0}(t, \rmeso,\theta, z) + \littleo{\delta^3}\, .
\end{alignat}
Expansions for the thermodynamic variables are anticipated as follows,
\begin{subequations}\label{eq:MesoExpansion}
\begin{alignat}{8}
  \label{Mes:PressureExpansion}
  &\p
  && = \pnull
  && + \epsonehalf^{2}\ptwo
  && + \epsonehalf^{4}\, \left(\pc\order{4} + \pc_{4}\right)
  && + \epsonehalf^{5}\, \left(\pc\order{5} + \pc_{5}\right)
  && + \littleo{\epsonehalf^{5}}\, ,
  \\
  \label{Mes:DensityExpansion}
  & \rho
  && = \rhonull
  && + \epsonehalf^{2}\rhotwo
  && + \epsonehalf^{4}\, \left( \rhoc\order{4} + \rhoc_4\right)
  && + \epsonehalf^{5}\, \left( \rhoc\order{5} + \rhoc_{5}\right)
  && + \littleo{\epsonehalf^{5}}\, ,
  \\
  \label{Mes:ThetaExpansion}
  & \Theta
  && = \Theta_0
  && + \epsonehalf^{2} \Thetatwo
  && + \epsonehalf^{4}\, \left( \Thetac\order{4}+ \Thetac_{4}\right)
  && + \epsonehalf^{5}\, \left( \Thetac\order{5}+ \Thetac_{5}\right)
  && + \littleo{\epsonehalf^{5}}\, ,
\end{alignat}
\end{subequations}
\citep[for plausibility arguments see][section 4.1.3]{PaeschkeEtAl2012}.
In \eq{eq:MesoExpansion}, the variables
$(p_0, p_2, \rho_0, \rho_2, \Theta_2)(z)$ describe the stationary
background ($\Theta_0$ is a constant), $(\pc_{i}, \rhoc_{i}, \Thetac_{i})(\tv, z)$, are higher-order
horizontal means, and
$\left(\pc\order{i}, \rhoc\order{i}, \Thetac\order{i}\right)(\tv, \rmeso, \theta, z)$
are the quantities of prime interest.

Note that, owing to the Fourier representation defined in \eq{eq:Fourier} this
notational convention ``overloads'' the subscript $(\cdot)_0$ with a double-meaning,
but the distinction should always be clear from the context.

The vortex centerline position is expanded as
\begin{equation}
\XC(\tv, z) =  \XC\order{0}(\tv, z) + \bigoh{\epsonehalf^1} \,.
\end{equation}
%


\section{Asymptotic analysis of the core structure evolution}
\label{sec:CoreStructureEvolution}

This section revisits the analysis of \citet{PaeschkeEtAl2012} for large vortex Rossby
numbers focusing on the evolution equation for the primary circulation.


\subsection{Asymptotic equation hierarchy for the vortex core}
\label{Asympe}

%
%

The governing equations transformed to the co-moving coordinates are provided in appendix~\ref{sec:TransformedEquations}. Inserting the expansion scheme from the previous section
we obtain
\begin{subequations}\label{eq:HorMomLeadingFirst}
\begin{alignat}{3}
\label{eq:HormomLeading}
\dss - \frac{(u_{\theta}\zero)^2}{\rmeso}
+ \frac{1}{\rho_0} \pderiv{\pc\order{4}}{\rmeso}
  & = 0\,,
    & \dss\qquad\pderiv{\pc\order{4}}{\theta}
      & = 0
        \\
\label{eq:HormomFirst}
- \frac{2 u_{\theta}\zero \uthet\order{1}}{\rmeso}
+\frac{1}{\rho_0} \pderiv{\pc\order{5}}{\rmeso}
- f_0 u\zero_{\theta}
  & = 0\,,
    & \qquad \pderiv{\pc\order{5}}{\theta}
      & = 0
\end{alignat}
\end{subequations}
from the horizontal momentum balance at leading and first order, respectively.
Each line in \eq{eq:HorMomLeadingFirst} displays the respective radial balance
first and the circumferential balance as the second equation.
We observe from the radial component in \eq{eq:HormomLeading} that the vortex
is in cyclostrophic balance to leading order which implies large vortex
Rossby number. The Coriolis effect enters only as a first-order perturbation
in the present regime as seen in the radial component of \eq{eq:HormomFirst}.
The pressure perturbations $p\order{4}, p\order{5}$ inherit the assumed axisymmetry
of $\uthet\order{0}, \uthet\order{1}$ thanks to the leading and first order
circumferential momentum balances in \eq{eq:HormomLeading} and \eq{eq:HormomFirst},
respectively.

The full second order horizontal momentum
equations are listed in appendix~\ref{sec:FullSecondOrderMomentum},
equations~\eq{eq:HormomSecondApp}, but for the rest of the paper we only need the
circumferential average of the circumferential component \eq{mom_theta_32}.
Letting $\psi_0 \equiv \frac{1}{2\pi}\int_{-\pi}^{\pi} \psi(\theta) \, d\theta$
denote the circumferential average of some $\theta$-dependent variable $\psi$ in
line with \eq{eq:Fourier}, we have
\begin{equation} \label{eq:MomThetaTwoAverage}
\pp{\uthet\zero}{t}
+ \wc\order{0}_0 \pp{\uthet\zero}{z}
+ u_{r,0}\order{2}\left( \pp{\uthet\zero}{\rmeso}
+ \frac{\uthet\zero}{\rmeso} \right)
- u_{r,*}\order{2}\pderiv{\uthet\zero}{\rmeso}
= 0 \,,
\end{equation}
where
\begin{equation}\label{eq:UrStarDef}
u_{r,*}\order{2} = \left(\wc\order{0} \verad \cdot\pderiv{\XC\order{0}}{z}\right)_0\,.
\end{equation}
The flow is hydrostatic up to third order, \ie,
$\pderiv{p_{i}}{z} = -\rho_i$ $(i = 1,...,4)$, whereas
\begin{equation} \label{vertmom2}
\frac{\partial \pc\order{4}}{\partial z} - \frac{\partial \XC\order{0}}{\partial z}
\cdot \verad \pderiv{\pc\order{4}}{\rmeso} \ = - \rhoc\order{4} \, .
\end{equation}
The leading and first order velocities are horizontal and axisymmetric according to
\eq{Meso:HorizontalVelocity}, \eq{Meso:VerticalVelocity} and thus divergence free.
The second order velocity is subject to an anelastic divergence constraint obtained
from the mass balance,
\begin{equation} \label{mass2}
\frac{\rho_0}{\rmeso}
\left(
  \pp{}{\rmeso}\left(\rmeso \urad\order{2}\right)
  + \pp{\uthet\order{2}}{\theta}
\right)
+ \frac{\partial}{\partial z} \left( \rho_0 \wc\order{0} \right)
- \frac{\partial \XC\order{0}}{\partial z} \cdot \vect{\hat{\nabla}}_\hor (\rho_0 \wc\order{0}) = 0\, .
\end{equation}
Similarly, the first non-trivial potential temperature transport equation reads
\begin{equation} \label{potentialtemp2}
\frac{\uthet\order{0}}{\rmeso} \pp{\Thetac\order{4}}{\theta}
+ \wc\order{0} \frac{d \Theta_2}{d z}
= Q_{\Theta}\order{0}\, ,
\end{equation}
and the equation of state relates the thermodynamic perturbation variables
through
\begin{equation}\label{state42}
\rhoc\order{4}
=
\rho_0
\left(
  \frac{\pc\order{4}}{\gamma\p_0} - \frac{\Thetac\order{4}}{\Theta_0}
\right)   \, .
\end{equation}
%


\subsection{Temporal evolution of the vortex structure}
\label{ssec:CoreEvolution}

\citet{PaeschkeEtAl2012} observed that with the aid of \eq{eq:HorMomLeadingFirst} and
\eq{eq:UrStarDef}--\eq{state42}, and given the vortex tilt, $\pptext{\XC\order{0}}{z}$,
as well as the diabatic source term, $\Qtheta\order{0}$, one may interpret
\eq{eq:MomThetaTwoAverage} as a closed evolution equation for the leading
circumferential velocity, $\uthet\order{0}$.

To corroborate this, we use the Fourier decomposition, \eq{eq:Fourier}, for $\wc\order{0}$
and  the representation of the centerline representation in polar coordinates from
\eq{eq:CenterlinePolar} to obtain
\begin{equation}\label{eq:UrStar}
u_{r,*}\order{2}
=
\left(\wc\order{0} \verad \cdot\pderiv{\XC\order{0}}{z} \right)_0
=
\frac{1}{2}
\left[
  \wc\order{0}_{11}\pp{\XCx\order{0}}{z} + \wc\order{0}_{12}\pp{\XCy\order{0}}{z}
\right]\, .
\end{equation}
Expressions for $\wc\order{0}_0$ and $\wc\order{0}_{1k}$ for $k = 1,2$
follow from the $\Theta$--transport equation in \eq{potentialtemp2},
\begin{equation} \label{w52}
\wc\order{0}_0 \dd{\Theta_2}{z} =  Q_{\Theta,0}\order{0}\, ,
\qquad
\wc\order{0}_{1k} \dd{\Theta_2}{z} = Q_{\Theta,1k}\order{0} -
      (-1)^k \frac{u_{\theta}\zero}{\rmeso} \Thetac\order{4}_{1[3-k]} \,.
\end{equation}
Since $\pc\order{4}$ is axisymmetric (see \eq{eq:HormomLeading}),
$\pc\order{4}_{1k} \equiv 0$ and the equation of state, \eq{state42},
yields $\Thetac\order{4}_{1k}/\Theta_0 = - \rhoc\order{4}_{1k}/\rho_0$.
With this information, the vertical momentum balance \eq{vertmom2} yields
\begin{equation}\label{vert11newtest2}
\frac{\Thetac_{11}\order{4}}{\Theta_0}
  = - \frac{\rhoc_{11}\order{4}}{\rho_0}
  =  - \frac{1}{\rho_0}\frac{ \partial \XCx\order{0}}{\partial z}
       \frac{\partial \pc\order{4}}{\partial \rmeso}\, ,
\qquad
\frac{\Thetac_{12}\order{4}}{\Theta_0}
  = - \frac{\rhoc_{12}\order{4}}{\rho_0}
  = - \frac{1}{\rho_0}\frac{ \partial \XCy\order{0}}{\partial z}
      \frac{\partial \pc\order{4}}{\partial \rmeso}\, .
\end{equation}
Using the cyclostrophic balance in \eq{eq:HormomLeading} to eliminate
$\pptext{\pc\order{4}}{\rmeso}$, and going back to \eq{w52} we obtain
expressions for the $\wc\order{0}_{1k}$ in terms of
$\uthet\order{0}$, $\pptext{\XC\order{0}}{z}$, and $\Qtheta\order{0}$,
\begin{equation} \label{w5211}
\wc\order{0}_{1k}\dd{\Theta_2}{z}
  = Q_{\Theta,1k}\order{0} -
         (-1)^k \Theta_0 \pp{X_{[3-k]}\order{0}}{z} \,
        \frac{(u_{\theta}^{(0)})^3}{\rmeso^2}
\qquad
\left(\rule{0pt}{10pt}k = 1,2
\right)\, ,
\end{equation}
where $\XCx\order{0}_1 \equiv \XCx\order{0}$ and $\XCx\order{0}_2 \equiv \XCy\order{0}$.
Upon insertion of this result in \eq{eq:UrStar}, the second term on the right cancels,
so that
\begin{equation}\label{eq:UrStarFinal}
u_{r,*}\order{2}
=
\frac{1}{2\, d\Theta_2/dz}
\left[
  Q_{\Theta,11}\order{0}\pp{\XCx\order{0}}{z} + Q_{\Theta,12}\order{0}\pp{\XCy\order{0}}{z}
\right]
\equiv
\frac{1}{2\, d\Theta_2/dz}
\vect{Q}\order{0}_{\Theta,1} \cdot \pderiv{\XC\order{0}}{z}
\, .
\end{equation}
Here we have interpreted the cosine and sine Fourier-1 components of $Q_{\Theta}\order{0}$
as the components of a heating dipole vector, $\vect{Q}_{\Theta}$, in the horizontal plane.

To find a corresponding expression for $u_{r,0}\order{2}$ (see the third term in \eq{eq:MomThetaTwoAverage}), consider the circumferential average of mass continuity, \eq{mass2}.
A brief calculation yields
\begin{equation} \label{radialvel22}
\pderiv{\left(\rmeso \rho_0 u_{r,0}\order{2}\right)}{\rmeso}
+ \frac{\partial\left(\rmeso \rho_0 \wc\order{0}_0 \right)}{\partial z}
- \frac{1}{2}
\left[
\pderiv{\XCx\order{0}}{z} \pderiv{(\rmeso \rho_0 \wc\order{0}_{11})}{\rmeso} + \pderiv{\XCy\order{0}}{z} \pderiv{(\rmeso \rho_0 \wc\order{0}_{12})}{\rmeso}
\right]
= 0
\end{equation}
or, equivalently,
\begin{equation} \label{anelastic}
\pderiv{\left(\rmeso \rho_0 \left[u_{r,0}\order{2} - u_{r,*}\order{2}\right]\right)}{\rmeso}
+ \frac{\partial\left(\rmeso \rho_0 \,\wc\order{0}_0 \right)}{\partial z}
= 0
\end{equation}
with $u_{r,*}\order{2}$ defined in \eq{eq:UrStar}. Exploiting \eq{w5211} in that definition
and integrating in $\rmeso$ requiring that $u_{r,0}\order{2}$ be finite at
$\rmeso=0$ we find
\begin{equation}\label{eq:urzeroone}
u_{r,0}\order{2} =
u\order{2}_{r,00} + u_{r,*}\order{2}\, ,
\end{equation}
where
\begin{equation}\label{eq:UradZeroZero}
u\order{2}_{r,00}
=
    - \frac{1}{\rmeso}
      \int_0^{\rmeso}
         \frac{r}{\rho_0}
         \pderiv{}{z}\left(\rho_0 \frac{Q_{\Theta,0}\order{0}}{{\rm d}\Theta_1/{\rm d} z}\right)
      \mathrm{d} r\, .
\end{equation}
With \eq{w52} (first equation), \eq{eq:UrStarFinal}, \eq{eq:urzeroone}, and \eq{eq:UradZeroZero}
we have now indeed expressed $w\order{0}_0, u_{r,0}\order{2}$, and $u_{r,*}\order{2}$ in
terms of $\uthet\order{0}, \pptext{\XC\order{0}}{z}$, and $Q_{\Theta}\order{0}$ as announced. In the
sequel, we may thus derive from \eq{eq:MomThetaTwoAverage}
how vortex tilt and diabatic heating affect the evolution of the primary circulation.

The results in this section match the corresponding result by \citet{PaeschkeEtAl2012} with
the Coriolis parameter $f_0$ set to zero. This corroborates our statement~(\ref{item:IntroA})
in the introduction that the vortex amplification/attenuation mechanism described
in their work does not depend on the vortex Rossby number being at most of order unity.


\section{Discussion of the asymmetric intensification/attenuation mechanism}
\label{sec:Discussion}


\subsection{The influence of asymmetric heating on the primary circulation}
\label{ssec:InfluenceOfAsymmetricHeating}

As elaborated in the previous section, \eq{eq:MomThetaTwoAverage} describes the
evolution of the primary circulation in response to external diabatic heating in the
present vortex flow regime. Aiming to separate the influence of heating asymmetries from
those of axisymmetric effects, we recall from \eq{eq:urzeroone} that the net
circumferentially averaged radial velocity is entirely a response to diabatic effects,
and that it consists of one part, $u\order{2}_{r,00}$, which, according to
\eq{eq:UradZeroZero} is induced by axisymmetric heating, and a second part,
$u_{r,*}\order{2}$, which, according to \eq{eq:UrStarFinal}, arises from first Fourier
mode asymmetric heating patterns. Using this decomposition in \eq{eq:MomThetaTwoAverage},
we rewrite the equation as
\begin{equation} \label{eq:MomThetaTwoAverageAsymmetriesSplitOff}
\pp{\uthet\zero}{t}
+ \wc\order{0}_0 \pp{\uthet\zero}{z}
+ u_{r,00}\order{2}\left( \pp{\uthet\zero}{\rmeso}
+ \frac{\uthet\zero}{\rmeso} \right)
= - u_{r,*}\order{2}\frac{\uthet\zero}{\rmeso}  \,,
\end{equation}
which is the large-Rossby version of equation \eq{eq:TempevoluthetaIntro} announced in the
introduction. In this equation, the left hand side captures the influence of the
axisymmetric dynamics and diabatic heating, whereas the right hand side covers all
effects due to the interaction of asymmetric heating and vortex tilt.


\subsection{Mechanics of vortex intensification by asymmetric heating of a tilted vortex}
\label{ssec:SpinUpMechanism}
\begin{figure}[htbp]
\begin{center}
\includegraphics[width=0.8\textwidth]{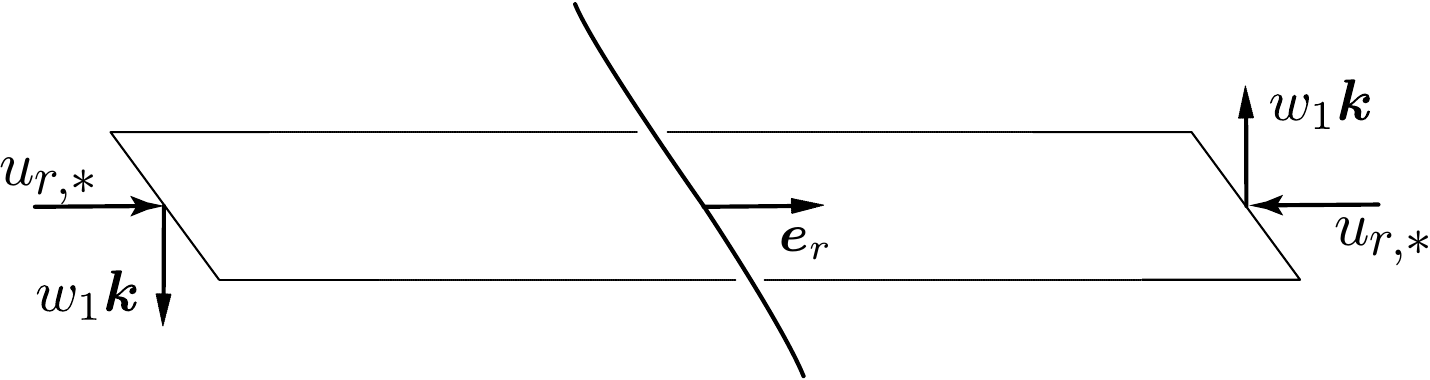}
\caption{Vertical cross section through a slant-cylindrical control volume of a tilted vortex.
Vertical mass transport induces by the vertical velocity dipole $\vect w_1$ through the boundary of the control volume is compensated by horizontal mass transport of opposite sign due to mass conservation.}
\label{fig:ApparentMassFlux}
\end{center}
\end{figure}

In the following section we analyze the leading-oder mass balance relations given in \eqref{radialvel22} and \eqref{anelastic}. We furthermore argue that $u_{r,*}\order{2}$ given in \eqref{eq:UrStarFinal} plays a crucial role in explaining the spin-up mechanism based on asymmetric diabatic heating. In this context we note that, according to \eqref{w5211}, the first order Fourier modes of the vertical velocity involve a contribution from diabatic heating (first term) and one due to the adiabatic dynamics (second term). It is only the contribution by diabatic heating that has an impact on $u_{r,*}\order{2}$ as seen in \eqref{eq:UrStarFinal}.

Depicting the situation of asymmetric heating anti-parallel to the tilt in figure~\ref{fig:ApparentMassFlux} we observe that the suitably arranged vertical motions can generate a mass flux through the boundary of the tilted disc control volume with the coordinate interval $(z, z+ \Delta z)$. Considering mass continuity in centerline-attached coordinates in \eqref{radialvel22}, we can identify the term in brackets as the axisymmetric mean of the vertical mass flux. Equation \eqref{anelastic} reveals that this expression is equal to a horizontal mass flux governed by $u_{r,*}\order{2}$. We therefore conclude, that the net vertical outflow in figure~\ref{fig:ApparentMassFlux} is compensated in the present balanced vortex situation by a net horizontal inflow to preserve continuity.

This gives an additional spin-up mechanism which exploits the vertical (tilted) structure of the vortex to gain angular momentum by moving air masses from larger radii to the center of the vortex. In contrast, the opposite orientation of diabatic Fourier-1 modes leads to an attenuation of the vortex by transporting angular momentum away from the center.
We therefore claim that by this mechanism it is possible to influence the overall strength of an atmospheric vortex as will be demonstrated in section \ref{sec:Simulations}.

This should settle the announcement of (item~\ref{item:IntroB}) in the introduction.


\subsection{Energy budget for the externally heated vortex}
\label{ssec:EnergyBudget}

Here we elaborate on how the asymmetric diabatic heating is transferred to kinetic
energy of the primary circulation in a tilted vortex.
This will be particularly useful in assessing the derived equations within the framework of Available Potential Energy \citep[APE,][]{Lorenz1955}.

To this end we multiply
\eq{eq:MomThetaTwoAverageAsymmetriesSplitOff} by $\rho_0 \rmeso\,\uthet\order{0}$,
use the $\theta$-averaged leading-order mass balance from \eq{anelastic} and recast
the advective terms in conservation form to obtain,
\begin{equation}\label{eq:KineticEnergyBalanceI}
\pderiv{}{t}\left(\rmeso \rho_0\frac{\uthet^2}{2}\right)
+ \pderiv{}{\rmeso}\left(\rmeso \rho_0 u_{r,00}\order{2} \frac{\uthet^2}{2}\right)
+ \pderiv{}{z}\left(\rmeso \rho_0 \wc_0 \frac{\uthet^2}{2}\right)
= - \rmeso\, u_{r,0}\order{2} \pderiv{p\order{4}}{\rmeso}\,.
\end{equation}
Here we have dropped the $\order{0}$ superscript on $\uthet\order{0}$ and
$w\order{0}$ to simplify
the notation, and we have used the cyclostrophic radial momentum balance from
\eq{eq:HorMomLeadingFirst} to introduce the pressure gradient on the right.

This reveals the change of kinetic energy (left hand side) to result from the work of
the pressure force due to the mean radial motion (right hand side). Some straightforward
but lengthy calculations, the details of which are given in
appendix~\ref{sec:KineticEnergyBudgetApp}, yield a direct relation of the kinetic
energy balance in \eq{eq:KineticEnergyBalanceI} to the Lorenz' theory of generation
of available potential energy (APE) by diabatic heating,
\begin{align}\label{eq:EKinBalance}
\left(\rmeso \ekin\right)_t
+ \left(\rmeso u_{r,00}\order{2} \hkin \right)_{\rmeso}
+ \left(\rmeso w_0\order{0} \hkin \right)_z
&= \frac{\rmeso\rho_0}{d\Theta_2/dz}\frac{1}{\Theta_0}
     \left[\Theta_0\order{4}Q_{\Theta,0}\order{0}
     +  \frac{1}{2}\vect{\Theta}_1\order{4} \cdot \vect{Q}_{\Theta,1}\order{0}
     \right]\,, \nonumber \\
&= \frac{r\rho_0}{N^2\Theta_0^2}
     \left(\Theta\order{4} \cdot Q_{\Theta}\order{0} \right)_0
\end{align}
where $\hkin = \ekin + p\order{4}$, and
$\left(\vect{\Theta},\vect{Q}_\Theta\right)_{1}
= \left(\Theta, Q_{\Theta}\right)_{12} \vi + \left(\Theta, Q_{\Theta}\right)_{11} \vj$
are the dipole vectors spanned by the first circumferential Fourier components of the
fourth order potential temperature perturbation, $\Theta\order{4}$, and of the diabatic
heating function, $Q_{\Theta}\order{0}$, respectively.

Equation \eqref{eq:EKinBalance} poses the differential form of kinetic energy balance.
To end up with an integral form as presented in \citet{Lorenz1955} we make use of the Gauss's theorem which allows us to drop the radial and vertical derivative assuming $u_{r,00}\order{2}$ and $w_0\order{2}$ vanish for sufficiently large $\rmeso$ and $z$ respectively.
To achieve this condition for $u_{r,00}\order{2}$ \eqref{eq:UradZeroZero} shows that we do not only need $Q_{\Theta,0}$ such that the integral converges for large radii but we need the integral to converge to zero.
When assuming a concentrated pattern of heat release with amplitude $\bigoh{10^{-4}\usk\kelvin\per\second}$ close to the vortex center (in the eyewall) over a surface of $\sim$ $10^4$ km${}^2$ it would need to be counteracted by contributions with opposite sign, \ie, cooling, but over a much larger surface of $\sim$ $10^6$ km${}^2$.
This simple scale approximation reveals cooling rates of $\bigoh{{\unit{0.1}{K/d}}}$ which is by an order of magnitude smaller than what is observed by radiative cooling \citep{KuhnLondon1969}.

For the total (integrated) kinetic energy $E_k$ we find
\begin{equation}
 \frac{dE_k}{dt} = 2\pi \int\limits_0^\infty \int\limits_0^\infty\frac{\rmeso \rho_0}{N^2{\Theta_0^2}} \left( \Theta\order{4} Q_\Theta\order{0} \right)_0 \, dr\,dz\,,
\end{equation}
On the one hand, \citet{Lorenz1955} balanced the kinetic energy with the \emph{conversion rate} from APE to kinetic energy ($C$) and the \emph{dissipation rate} ($D$) where the latter is neglected here.
On the other hand, the expression on the right-hand side coincides with the \emph{generation rate} ($G$) of APE (see appendix \ref{sec:APE} for details).
Therefore, no APE accumulates in the present flow regime as it is directly converted to kinetic energy (at leading order).
This is the result of the timescale used in the asymptotic analysis as conversion between APE and kinetic energy is accomplished by the advective and pressure-velocity fluxes on faster timescales.

In line with \citet{Lorenz1955,Lorenz1967} and announced in the introduction in item \ref{item:IntroC} this result shows that positive correlations of temperature perturbation and diabatic heat release lead to the increase of kinetic energy.
The precise form of the right hand
side of \eq{eq:KineticEnergyBudget} as announced in the introduction
(item~\ref{item:IntroC}) is obtained from \eq{eq:EKinBalance} by realizing that
$(1/\Theta_0) d\Theta_2/dz$ is the dimensionless representation of $N^2$, the square
of the Brunt-V\"ais\"al\"a frequency, and that the constant $\Theta_0$ is the
leading-order dimensionless background potential temperature
$\Thetabar = \rfr{T} (\Theta_0 + \littleoh{1})$.

\citet{NolanEtAl2007}, extending prior similar studies, investigate the influence
of asymmetric diabatic heating on vortex intensification on the basis of a linearized
anelastic model that includes a radially varying base state and baroclinic primary
circulation. Their central conclusions are that (i) asymmetric heating patterns
quite generally tend to attenuate a vortex, that (ii) there are situations in which
they can induce amplification, but in these cases their influence is (iii) generally
rather weak. In fact, they state in their section e: ``... purely asymmetric heating
generally leads to vortex weakening, usually in terms of the symmetric energy, and
always in terms of the low-level wind.''. Equation \eq{eq:EKinBalance} shows, in contrast, that purely
asymmetric heating in a tilted vortex can intensify or attenuate a vortex depending
on the arrangement of the heating pattern relative to the tilt, and that the efficiencies
of symmetric and asymmetric heating in generating kinetic energy are of the same order
in the asymptotics as claimed in (item~\ref{item:IntroC}) of the introduction.


\subsection{Diabatic forcing of the centerline motion}
\label{ssec:HeuristicsCenterlineMotion}
Together with the previous discussion we want to highlight some aspects of the effects of asymmetric heating on the vortex centerline motion.
Examinations in appendix \ref{sec:CenterlineMotion} of the constituents of the centerline equation reveal that the term $\vect k \times \vPsi$ splits into an adiabatic and a diabatic contribution due to the linear dependency of $\vPsi$ on the vertical velocity dipole $\vect w_1$ and $\vect w_1$ being composed of an adiabatic and a diabatic contribution (see \eqref{w5211}).
In particular, we find that the adiabatic expression is of the same form as the diabatic one but evaluated with the adiabatic vertical velocity:
\begin{equation}
  \vect w_{1,\text{ad}} = - \begin{pmatrix}
                              0 & 1 \\
                              -1 & 0
                            \end{pmatrix}
    \frac{1}{\Theta'_2} W \partial_z \vect X
    = \hat R_{-\pi/2} \frac{1}{\Theta_2'} W \partial_z \vect X\,,
  \label{eq:w_ad}
 \end{equation}
with
\begin{equation}
  W = \frac{u_\theta}{r} \left( \frac{u_\theta^2}{r} + f u_\theta \right)\,,
\end{equation}
and the rotation matrix $\hat R_{\theta_0}$.
Inserting \eqref{eq:w_ad} into eq. \eqref{eq:Psi_components} $\vPsi$ results in a linear differential operation on $\vect X$ which, interpreted as Hamiltonian of a (complex-valued) Schrödinger equation \eqref{eq:CenterlineSchroedinger}, leads to a purely real spectrum, \ie, to precession of the centerline in the complex ($x$-$y$) plane.

The diabatic motion of the centerline on the other hand results from inserting some non-trivial $\vect w_{1,\text{dia}}$ into $\vPsi$ which in our case shall be a rotated version of \eqref{eq:w_ad}:
\begin{equation}
  \vect w_{1,\text{dia}} = \hat R_{\theta_0} \frac{1}{\Theta_2'} W \partial_z \vect X\,.
  \label{eq:w_dia}
 \end{equation}
Here $\theta_0$ is the relative orientation of the diabatic vertical velocity dipole relative to the tilt.
Clearly, $\vect w_{1,\text{dia}}$ coincides with $\vect w_{1,\text{ad}}$ for $\theta_0=-\pi/2$.
Therefore, such a diabatic heating vertical velocity pattern will result in an additional contribution to the centerline motion of the same orientation and magnitude as the adiabatic contribution.
By numerical experiments we observed that the adiabatic vertical velocity dipole, oriented $-\pi/2$ relative to the tilt leads to a centerline motion in the direction $+\pi/2$.

In contrast by varying $\theta_0$ we expect a diabatic centerline forcing which is oriented $-\theta_0$ relative to the tilt.
Therefore, for our formulation of diabatic heating the rotation angle $\theta_0$ determines whether the diabatic forcing leads to acceleration ($\theta_0=-\pi/2$) or deceleration ($\theta_0=\pi/2$) of the centerline precession or to increasing ($\theta_0=\pi$) or decreasing the tilt of the centerline.%
\footnote{Note that the resulting linear operator governing the diabatic centerline motion effectively is the rotated version of the adiabatic operator (multiplied with complex number on the unit circle), \ie, its spectrum is rotated in complex plane as $\vect w_1$ is rotated.
Real parts of the spectrum lead to precession while imaginary parts lead to growth/damping of the centerline amplitude.}

In the further course of diabatic experiments we make use of our findings and specify heating patterns of the form
\begin{equation}
  \vect Q_{1,\theta_0} := \Theta_2' \vect w_{1,\text{dia}} = \Rotmat_{\theta_0} W \partial_z \vect X.
  \label{eq:Q1_theta_0}
\end{equation}


\section{Comparison between asymptotic model and three-dimensional simulations}
\label{sec:Simulations}

In the course of this paper we presented the derivation of a system of PDEs governing the leading-order dynamics of a tilted tropical cyclone in weak cyclostrophic regime (eqs. \eqref{eq:TempevoluthetaIntro} and \eqref{eq:TempevolutionCenterlineIntro}).

The aim of this section is twofold:
First, we want to validate the reduced model equations against the three-dimensional Euler equations, \ie, first principle equations \eqref{eq:CompressibleFlowEquationsDimless}, by solving both sets of equations in a suitable numerical framework.
To this end, we follow the work of \citet{Papke2017} who presented the first results on this scenario.
The second goal is to highlight principal mechanisms that are activated by purely asymmetric heating.
Therefore, after analyzing the adiabatic dynamics of an initially tilted vortex, we continue by constructing a prototypical asymmetric diabatic heating pattern which will be imposed on the vortex under different angles relative to the tilt.
We will refer to these experiments as \emph{adiabatic} (reference simulation), \emph{stagnation}, \emph{intensification} and \emph{attenuation} according to their influence on tilt and centerline.

The quasi-two-dimensional equations \eqref{eq:TempevoluthetaIntro} and \eqref{eq:TempevolutionCenterlineIntro} are solved by a combination of appropriate numerical methods details of which are presented in section \ref{sec:NumericsAsymptotics}.
For the three-dimensional simulations the general purpose atmospheric flow model EULAG \citep[see, \textit{e.g.},][]{PrusaEtAl08,SmolarkiewiczEtAl2014} provides efficient integration strategies for equations \eqref{eq:CompressibleFlowEquationsDimless}, and its compressible model was used. These two alternative representations of the tilted vortex flows will be referred to as \emph{asymptotic} and \emph{three-dimensional} simulations.

Note that, although we worked out the scaling in the cyclostrophic regime ($\Ro = \bigoh{1/\delta}$), initial data in the following sections will be in the gradient-wind regime ($\Ro = \bigoh{1}$).\footnote{Experiments have shown substantial damping of the centerline tilt amplitude when initializing the vortex in the cyclostrophic regime.
Furthermore, the special formulation of asymmetric diabatic heat release sensitively depends on the magnitude of tangential velocity causing numerical instabilities in the cyclostrophic regime.}
Therefore, in the numerical treatment terms involving the Coriolis parameter $f$, even if reasonably small, will not be neglected.

\subsection{Numerical setting and initial data}
\label{ssec:InitialData}

With the asymptotic analysis above we demonstrated that a tilted vortex evolves at leading-order on a time scale comparable to the synoptic time scale.
Higher-order dynamics occurs in the presence of initial perturbations (excitation of higher-order asymptotic modes) on faster time scales.
However, we are interested in the leading-order effects only, hence we construct initial data to closely reproduce the leading-order symmetries imposed by the asymptotic analysis of section \ref{sec:MesoScaleAnalysis} allowing to solve for solutions on the slowest varying manifold.

In the case of an adiabatic vortex the tangential velocity equation \eqref{eq:TempevoluthetaIntro} is trivially stationary,
\begin{equation}\label{eq:uth_noQ}
  \pderiv{}{t} u_{\theta}\order{0}=0\,,
\end{equation}
and the centerline equation \eqref{eq:TempevolutionCenterlineIntro} becomes a Schrödinger-type equation,
\begin{equation}
 i \partial_t X = \hat H X\,,
 \label{eq:CenterlineSchroedinger}
\end{equation}
with a Hamiltonian $\hat H$ depending on the tangential velocity $u_\theta(r,z)$ and the $z$-dependent background profiles $\bar\rho$, $\bar p$, $\bar\Theta$.
Note, that we have introduced $X := (\vect X \cdot \vect i) +  i (\vect X \cdot \vect j)$ as the complex representation of $\vect X$
(see appendix \ref{sec:CenterlineMotion}).

With suitable boundary conditions the governing Hamiltonian takes the form of a Sturm-Liouville operator and therefore exhibits a real, discrete spectrum \citep{Teschl2012}, which sets the precession frequency of each eigenmode.
The first non-trivial eigenvalue corresponds to the slowest varying solution and posses a \emph{cosine-like} eigenfunction, \ie, the simplest tilted solution.\footnote{Details on that are skipped but can easily be validated by numerical means.}

\begin{figure}[htbp]
\centering
\tikz{%
 \node[] (A) at (0,0) {\includegraphics[height=0.37\textwidth]{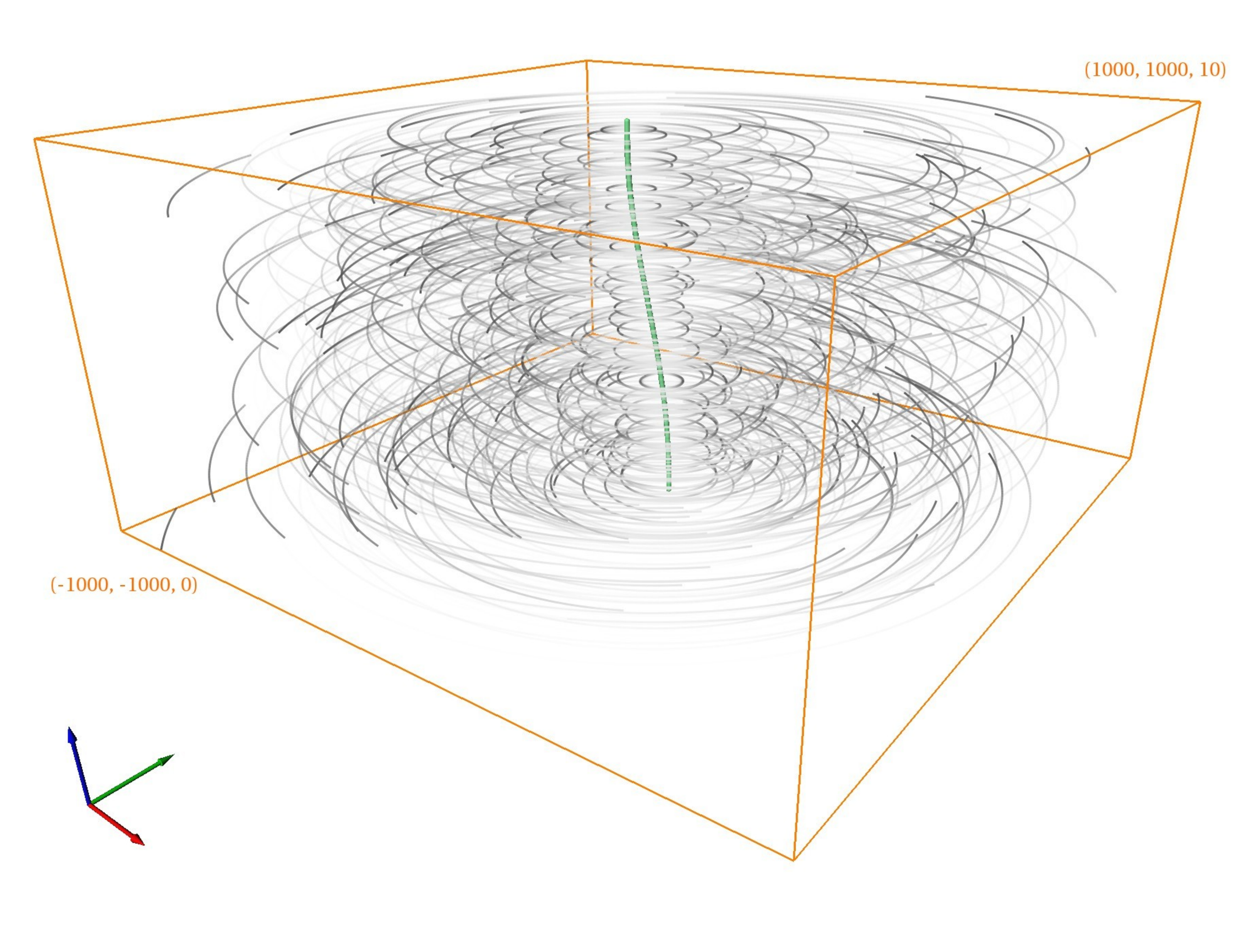}};
 \node[xshift=0.9cm, yshift=0.6cm] () at (A.south west) {\scriptsize$x$};
 \node[xshift=1cm, yshift=1cm] () at (A.south west) {\scriptsize$y$};
 \node[xshift=0.45cm, yshift=1.25cm] () at (A.south west) {\scriptsize$z$};
 \node[anchor=north west, yshift=-0.5cm] (B) at (A.north east) {\includegraphics[height=0.3\textwidth]{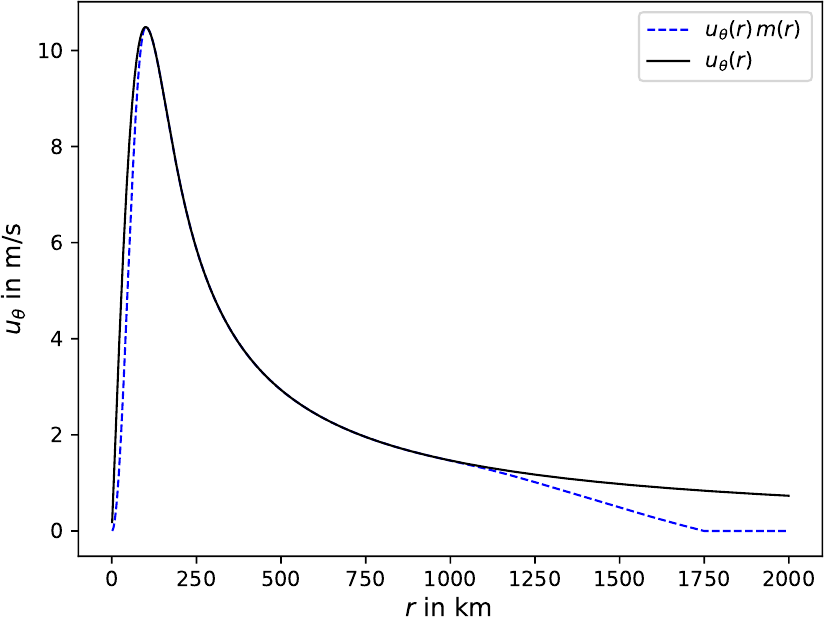}};
}
\caption{Initial setup of a Gaussian radial profile of vertical vorticity along a tilted centerline (left) and corresponding radial profile of circumferential velocity (right).
On the left, the inner simulation domain ($1000\usk\kilo\meter \times 1000\usk\kilo\meter \times 10 \usk\kilo\meter$) is displayed showing the initial centerline (green) along with the velocity field depicted by streamlines (gray).
On the right, the solid line corresponds to the unmodified profile \eqref{eq:initVelOrig} and the dashed line to the profile after applying the mollifier (see \eqref{eq:mollifier}).
}
\label{fig:uthProfile}
\end{figure}

As tilt is crucial for coupling asymmetric diabatic heating modes to the leading-order vortex dynamics, we prescribe an initially barotropic tangential velocity profile $u_\theta(r)$, corresponding to a Gaussian vorticity profile, $q(r)$,
\begin{equation}\label{eq:initVelOrig}
 \uthet = q_m \frac{1 - e^{-\sigma^2 r^2}}{2 \sigma^2 r}\,,
\end{equation}
where the radial vorticity profile reads
\begin{equation}
q(r) = q_m e^{-\sigma^2 r^2},
\end{equation}
and choose the first non-trivial eigenfunction of the corresponding Hamiltonian to define the initial centerline geometry, scaled to a displacement of 160 km, see figure~\ref{fig:uthProfile}.

For the sake of complying with horizontal boundary conditions imposed by EULAG and to avoid reflective features near the boundary of the computational domain, the initial data for the three-dimensional simulations are smoothly transitioned to zero at some finite radius by applying a mollifier:
\begin{equation}\label{eq:mollifier}
  m(r) =
  \begin{cases}
  \sin^2\left( \frac{\pi}{2} \frac{r}{r_0} \right) &, \quad r < r_0\\
  1 & ,\quad r_0 < r < r_1 \\
  \cos^2\left( \frac{\pi}{2} \frac{r-r_1}{\r_1-r_\infty} \right) &
    ,\quad r_1 < r < r_\infty\\
  0 & ,\quad r > r_\infty
  \end{cases}
\end{equation}
$r_1$ = 1 250 \kilo\meter~and $r_\infty$ = 1 750 \kilo\meter~are the radii where the mollifier starts and where it reaches full-suppression. In addition to that the profile within $r_0$ = 100 \kilo\meter~is adjusted to preserve differentiability at the origin.

As mentioned earlier, the solution depends on the background state of the atmosphere which is determined by the potential temperature profile
\begin{equation}
  \bar\Theta(z) = \rfr{T} \, \exp\left(\frac{\rfr{N}^2}{g} \, z\right)\,.
  \label{eq:expPotTemp}
\end{equation}

In the three-dimensional case the vortex is embedded into a domain of 4000 \kilo\meter~extent in
both horizontal directions, 10 \kilo\meter~in the vertical, and a damping layer surrounds the domain
near the horizontal boundaries to suppress gravity waves emerging from the inner core and to keep them from reflecting.

\begin{figure}[htbp]
 \centering
 \includegraphics[width=0.5\textwidth]{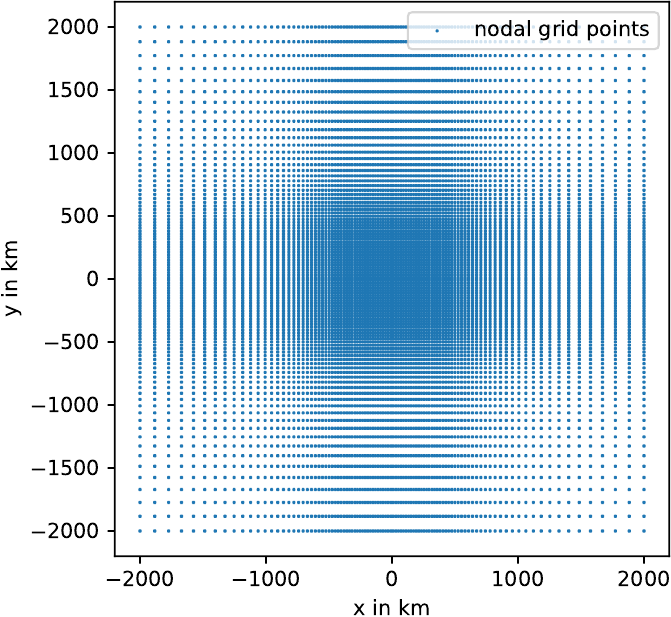}
 \caption{Horizontal structure of stretched grid used for solving the three-dimensional Euler equation with EULAG.}
 \label{fig:grid}
\end{figure}

Along the structural properties of the tropical cyclone, i.e., an inner core and smooth transition to the quasi-geostrophic far field, we use the static mesh refinement capability of EULAG \citep{PrusaEtAl08} and map equidistant coordinates onto a grid focused at the inner core.
The actual mapping is accomplished by
\begin{equation}
 \begin{pmatrix}
  x_p \\ y_p
 \end{pmatrix}
 = c_1
  \begin{pmatrix}
  x_c \\ y_c
 \end{pmatrix}
  +
  c_2
  \begin{pmatrix}
   x_c^\alpha \\ y_c^\alpha
  \end{pmatrix}\,,
\end{equation}
where $\alpha = 5$, $c_{1,2} = 1/2$.
$(x,y)_{p,c}$ are normalized coordinates on the domain $[-1,1]^2$.
Figure~\ref{fig:grid} demonstrates how the horizontal grid is focused towards the center of the computational domain.

The asymptotic equations are solved on a regular equidistant tilted polar grid on the domain $(r,z) \in [0, 1.12] \times [0, 12.5]$ in dimensionless units covering roughly a tilted cylinder of 1000 \kilo\meter~ around the centerline and the full vertical extent.

In the further course, we will compute the diabatic heating from eq. \eqref{eq:Q1_theta_0} which involves the reconstruction of both, the current centerline position and circumferentially averaged tangential velocity.
For the first, we compute the center of mass of the vorticity field at each horizontal level:
\begin{equation}
  \vect X = \int \vect k \cdot (\nabla \times \vect u_{||}) \vect x \, dx dy
  \label{eq:centerline_reconstruction}
\end{equation}
We follow the ideas of \citet{NguyenEtAl2014} and the implementation outlined in \citet{Papke2017}.

The circumferential mean of $u_\theta$ then is computed by the Biot-Savart integral
\begin{equation}
  u_{\theta,0} = \frac{\Gamma(r,z)}{2\pi r} = \frac{1}{2 \pi r} \int\limits_{B_r(\vect X)} \vect k \cdot (\nabla \times \vect u_{||}) \, dx dy\,,
\end{equation}
where $B_r(\vect X) = \{(\vect x_{||} \in \mathbb R^2 | (\vect x_{||} - \vect X)^2 < r^2 \}$ denotes the circular domain centered at $\vect X$ with radius $r$.

\subsection{Results and discussion}
\label{ssec:SimResDis}

In the following subsection we will present results of numerical simulations solving either the full three-dimensional Euler equations \eqref{eq:CompressibleFlowEquationsDimless} or the reduced asymptotic equation \eqref{eq:TempevoluthetaIntro} for the primary circulation velocity $u_\theta\order{0}$, and the centerline evolution \eqref{eq:TempevolutionCenterlineIntro} explained in detail in App.~\ref{sec:CenterlineMotion} and following.

\subsubsection{Adiabatic vortex}
\label{ssec:adiabatic}

As a reference for the following experiments we first investigate the dynamics of a tilted, adiabatic vortex.
In subsection \ref{ssec:InitialData} we constructed initial data to follow the first non-trivial eigenmode of the governing (adiabatic) Hamiltonian and we found stationarity (eq. \eqref{eq:uth_noQ}) of the mean tangential wind.
Hence, from the structure of eq. \eqref{eq:CenterlineSchroedinger} we expect undamped precession of the eigenfunction.

\begin{figure}[htbp]
 \centering
 \includegraphics[width=0.49\textwidth]{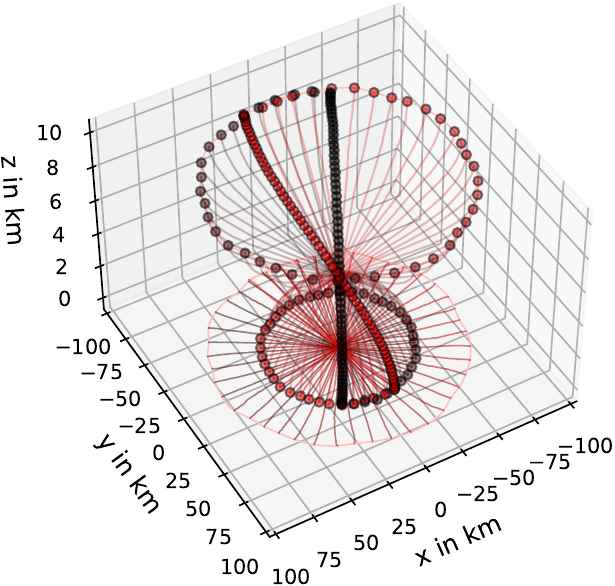}
 \hfil
 \includegraphics[width=0.49\textwidth]{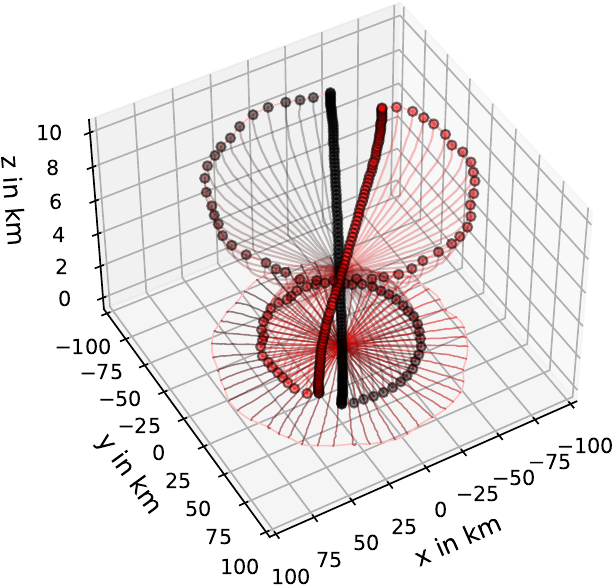}
 \caption{Time evolution of centerline for an adiabatic vortex for 6 days.
          Shown are the results of the asymptotic (left) and the three-dimensional simulation (right).
          In both panels coloring of the centerline transits from black to red indicating counterclockwise precession.
          Initial ($t=0$) and final ($t=6\usk\days$) are highlighted with circle markers along the vertical extent of the centerline.
          Endpoints (top and bottom) are marked with circles of the respective color for each time step.
          Additionally, the centerline is projected onto the bottom surface.
          }
 \label{fig:CenterlineMotion}
\end{figure}

Figure~\ref{fig:CenterlineMotion} compares the results of both adiabatic simulations, three-dimensional and asymptotic, with initial data as discussed.
Though exhibiting small-scale oscillation and damping, the three-dimensional simulation (fig. \ref{fig:CenterlineMotion}, right panel) compares well with its asymptotic analog.
Time scales of one precession are 5.5 days for the asymptotic and 6.5 days for the three-dimensional simulation.
This difference leads to a deviation of the final positions of the centerline but given that the effective expansion parameter $\delta = \sqrt{\eps} \sim 1/3$, this is well within the error bounds of the leading order solution.

\begin{figure}[htbp]
\centering
\includegraphics[height=0.37\textwidth]{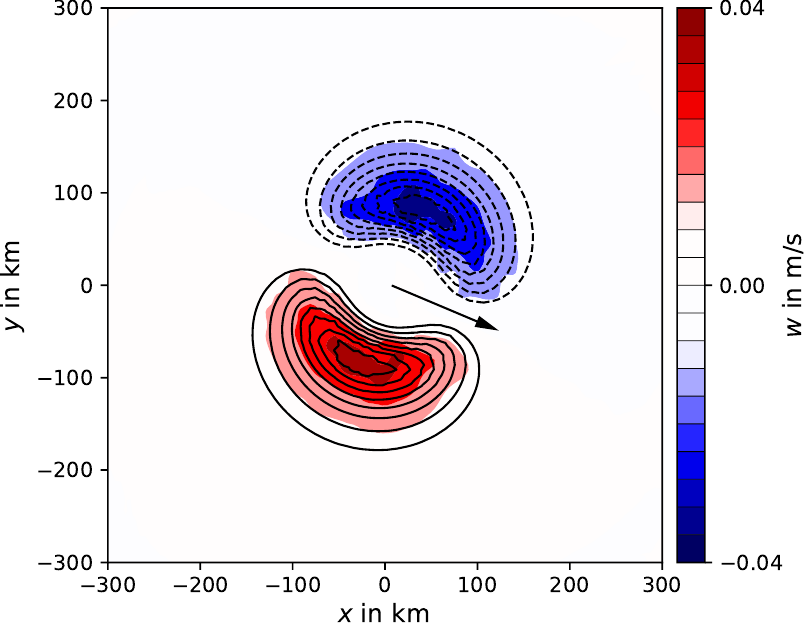}
\hfill
\includegraphics[height=0.37\textwidth]{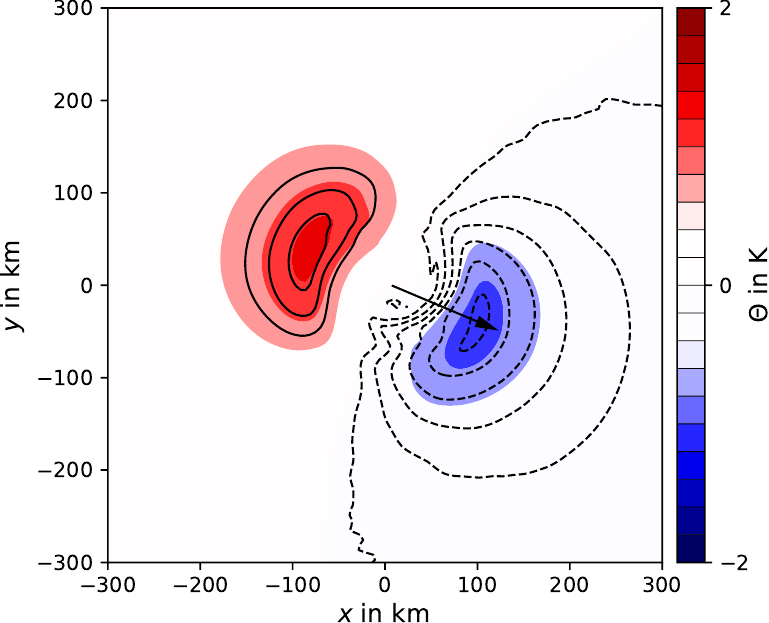}
\caption{Horizontal slices of vertical velocity (left) and potential temperature (right).
Color shades depict the numerical results at $5000\usk\meter$ height and $1.5\usk\days$ with values ranging from $-0.04 \usk\meter\per\second$ to $0.04 \usk\meter\per\second$ in steps of $0.004 \usk\meter\per\second$ (left) and from $-2\usk$K to $2\usk$K in steps of $0.2\usk$K with positive (red) and negative (blue) sign.
Contour lines correspond to the asymptotic prediction (solid for positive and dashed for negative sign).
The arrow indicates position and tilt direction of the centerline.}
\label{fig:wThAdiab}
\end{figure}

The asymptotic analysis revealed non-trivial leading-order balances for $w$ and the thermodynamic quantities according to tilt, gradient-wind (cyclostrophic) and hydrostatic balance.
In figure~\ref{fig:wThAdiab} both, $w$ and $\tilde\Theta$ are visualized by representative horizontal slice at 5000\usk\meter~ at $t = 6.5\usk\days$ comparing asymptotic values (black contours) with the 3D numerical simulation results (color-coded).
The tilt vector $\partial_z \vect X$ is indicated by an arrow.
Qualitative similarities are rather apparent, while deviations are again well within the asymptotic truncation order $\bigoh{\delta} \sim 1/3$.
Both figures demonstrate the alignment of dipolar perturbations relative to the tilt as $\vect w_1$ and $\vect \Theta'_1$ are rotated by -90\degree~ and 180\degree, respectively.
This is in agreement with the findings of \citet{Jones1995}.

\subsubsection{Stagnation}
\label{ssec:stagnation}

The stagnation test follows the idea that the choice of $\theta_0 = \pi/2$ in \eqref{eq:Q1_theta_0} leads to deceleration of the centerline precession by canceling the term $\vPsi$ in \eqref{eq:TempevolutionCenterlineIntro}.

Furthermore, it has no impact on the leading-order tangential velocity as we immediately see when neglecting symmetric vertical motions due to symmetric diabatic heat release resulting in the tangential velocity time evolution
\begin{equation}
  \pderiv{u_\theta}{t} = -u_{r,*}\left(\frac{u_\theta}{r} + f \right)
  \label{eq:TempEvoluUthetaAsymm}
\end{equation}
and realizing that $u_{r.*}$ is the projection of the diabatic heat release onto the tilt vector,
\begin{equation}
  u_{r,*} = \frac{1}{2} \left( \partial_z \vect X \cdot \vect Q_{1,\pi/2} \right)
          = \frac{1}{2} \left( \partial_z \vect X \cdot W R_{\pi/2} \partial_z \vect X \right)\,,
  \label{eq:urstar0}
\end{equation}
which vanishes due to orthogonality.

\begin{figure}[htbp]
\centering
\includegraphics[height=0.37\textwidth]{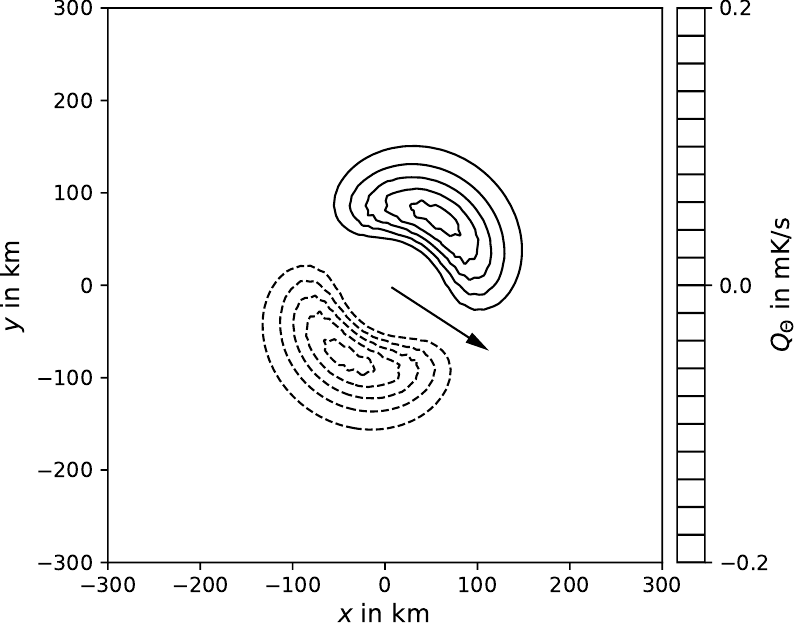}
\hfill
\includegraphics[height=0.37\textwidth]{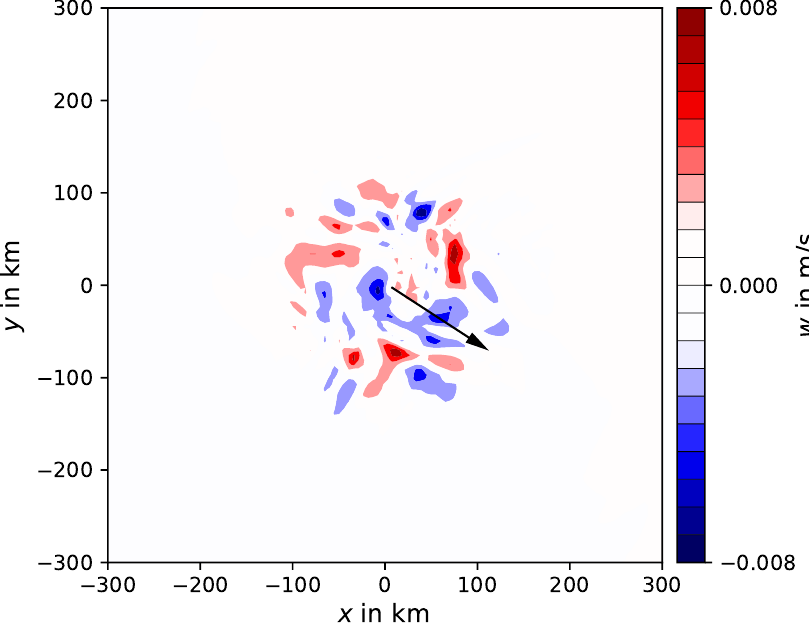}
\caption{Asymmetric heating pattern in stagnating configuration with orientation of $\theta_0 = \pi/2$ relative to the tilt (left) and residual vertical velocity (right) on a horizontal slice at 5000\usk\meter~and 1.5 days.
 Left: Contour lines represent absolute values from $0.02 \usk\mathrm{mK}/\second$ to $0.12 \usk\mathrm{mK}/\second$ in steps of $0.02 \usk\mathrm{mK}/\second$.
 Right: Color shades represent absolute values up to $0.008\usk\meter\per\second$ with positive (red) and negative (blue) sign.}
\label{fig:w_stag}
\end{figure}

\begin{figure}[htbp]
\centering
\includegraphics[height=0.45\textwidth]{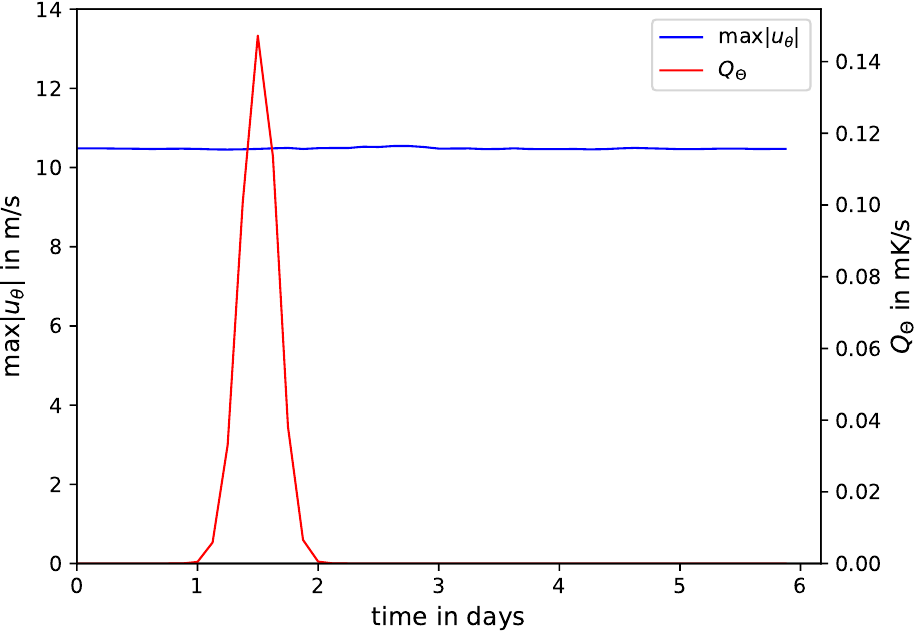}
\caption{Time series of maxima of three-dimensional tangential wind (blue) and diabatic heating pattern \eqref{eq:Q1_theta_0} (red) in the stagnation configuration.
As predicted by the theory no significant impact on the horizontal velocity is observed.}
\label{fig:MWS_stag}
\end{figure}

By construction, inserting the heating $\vect Q_{1, \pi/2}$ into \eqref{w5211} satisfies $\vect w_1 \equiv 0$ up to leading order.
Figure~\ref{fig:w_stag} shows the expected behavior for the three-dimensional simulations.
$\vect Q_{1, \pi/2}$ attenuates vertical velocity by a factor of $\sim\delta$, \ie, canceling the leading-order expansion mode of $w$.
The residual, depicted in the right panel may refer to higher-order expansion modes.
Furthermore, the leading-order tangential velocity is not affected by asymmetric heating of that orientation.
Figure~\ref{fig:MWS_stag} presents both the time series of maximum tangential wind (blue) and of maximum heating (red).
Only small variations of the tangential velocity are apparent and, as we will see in the next subsection, this changes substantially when we alter the orientation of $\vect Q_{1, \theta_0}$.

Experiments (not shown here) revealed instabilities caused by small perturbations due to discretization errors:
The diabatic heating will exhibit large amplitudes where the local tilt $\partial_z \vect X$ (being result of the reconstruction \eqref{eq:centerline_reconstruction}) is large.
This affects the local velocity and as a consequence the spectral properties of the Hamiltonian of eq. \eqref{eq:CenterlineSchroedinger} (projecting onto higher frequency modes) increasing small-scale oscillatory features of the centerline $\vect X$.
Hence, we need make sure to maintain a certain regularity of $\vect X$ to avoid obscuring the effect under consideration by triggering this feedback loop.
We achieve this by restricting the heating to a concentrated pulse by applying a time-dependent amplitude factor of the shape
\begin{equation}
  f(t) = a \exp\left( \frac{(t-b)^2}{2c^2} \right)
  \label{eq:blending}
\end{equation}
For the current setting $a=1$, $b=5.5 \mbox{ days}$ and $c=\frac{1 \mbox{day}}{2\sqrt{2 \ln 300}}$.

\begin{figure}[htbp]
 \centering
 \includegraphics[width=0.49\textwidth]{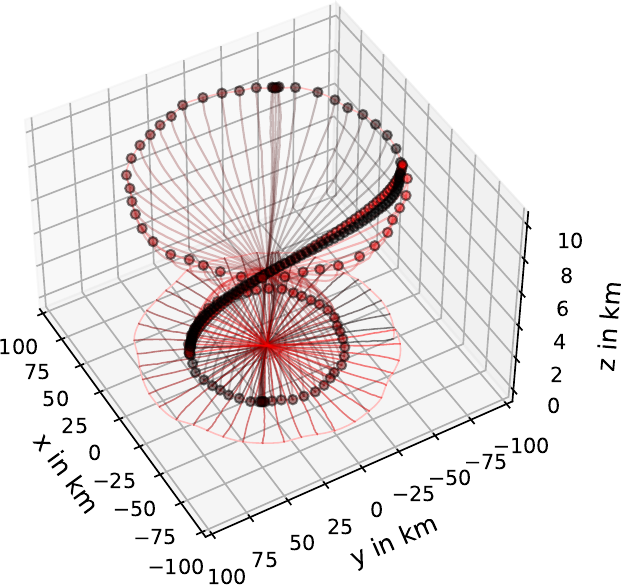}
 \hfill
 \includegraphics[width=0.49\textwidth]{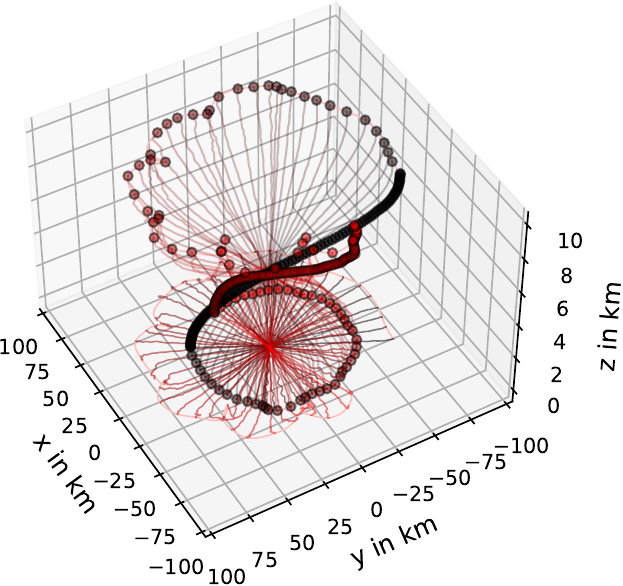}
 \caption{Same as figure \ref{fig:CenterlineMotion} but for a vortex under the influence diabatic heating (stagnation case with $\theta_0 = \pi/2$).
 Left: Asymptotic solution for 6 days with diabatic heating applied between day 1 and 2
 slowing down the precession speed.
 Right: Three-dimensional reference simulation with same heating parameters.}
 \label{fig:CenterlineMotionStagnation}
\end{figure}

The heating distribution $\vect Q_{1,\pi/2}$ constructed in such a way satisfies $u_{r,*}$ to vanish (cf. \eqref{eq:urstar0}) and additionally, by canceling $\vect w_1$, $\vPsi$ is canceled in the centerline equation of motion \eqref{eq:TempevolutionCenterlineIntro}. As discussed in the introduction of section \ref{sec:Simulations}, we are initially in the regime of gradient-wind balance.
Hence $\vect M_1$ remains the only contribution to the centerline equation of motion.
Numerical evaluations show that the amplitude of $\vect M_1$ is about 1/6 of the amplitude of the $\vPsi$ in the adiabatic case which is why we observe a significant slow down of the centerline precession in the asymptotic case in figure~\ref{fig:CenterlineMotionStagnation}, left panel, during the heating time interval (cf. fig \ref{fig:CenterlineMotion} for reference).
Although not as prominently as with the asymptotic simulation due to the aforementioned discretization errors, the centerline also slows down in the three-dimensional simulation.
I becomes also present that due to heating the shape of the centerline becomes superimposed by higher-frequent features that are excited due to the slight misalignment of the diabatic heating dipole.

\subsubsection{Intensification}
\label{ssec:intensification}

In subsection \ref{ssec:HeuristicsCenterlineMotion} we constructed a dipolar heating pattern and by aligning its orientation $\pi/2$ relative to the centerline tilt we found stagnation of the centerline precession as well as suppression of the vortex-induced vertical velocity dipole.
However, what is the influence of this heating dipole when oriented with $\theta_0 = \pi$ relative to the tilt?
$\theta_0=\pi$ determines the sign of \eqref{eq:urstar0} negative and as a consequence the right-hand side of \eqref{eq:TempEvoluUthetaAsymm} positive, \ie, leading to intensification.
To examine the effect of the orientation of a dipolar heating pattern without altering its amplitude by the dependency of $\vect Q_{1,\theta_0}$ on $u_\theta$ and $\partial_z \vect X$ we fix $u_\theta$ and $|\pderiv{\vect X}{z}|$ in \eqref{eq:Q1_theta_0} to values at $t=5\usk\days$ but still keep track of the orientation of $\pderiv{\vect X}{z}$.

\begin{figure}[htbp]
\centering
\includegraphics[height=0.325\textwidth]{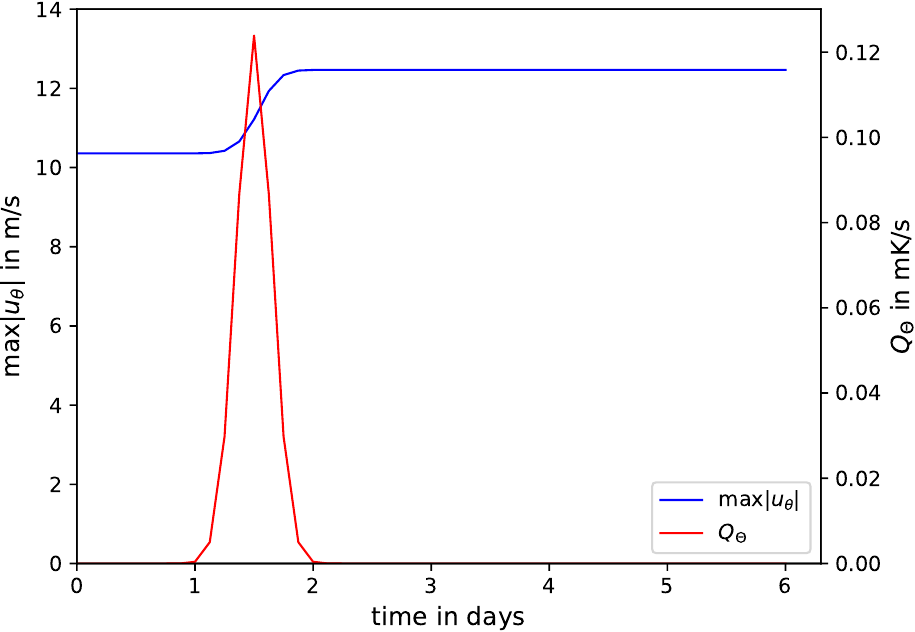}
\hfill
\includegraphics[height=0.325\textwidth]{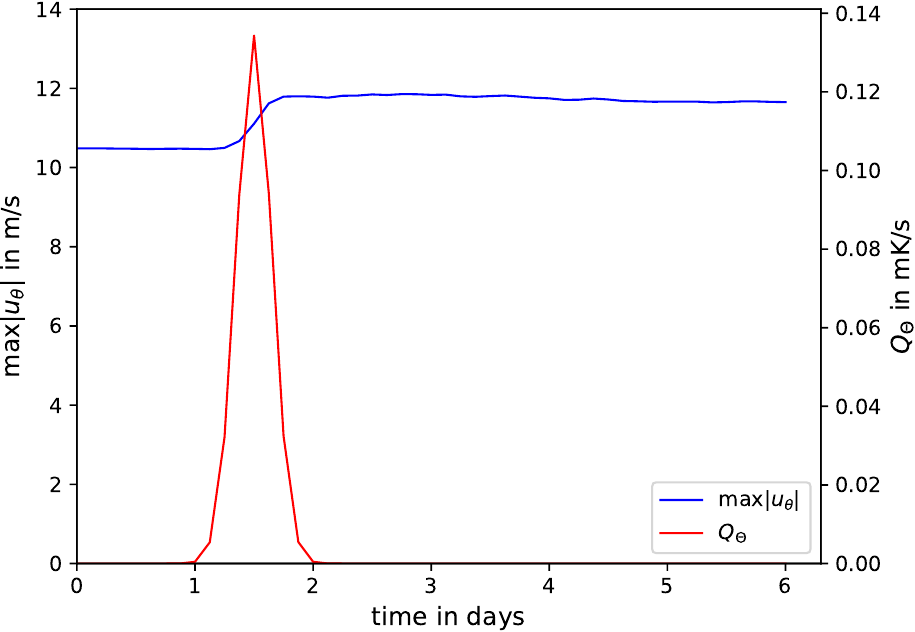}
\caption{Same as figure \ref{fig:MWS_stag} with $\theta_0=\pi$ (intensifying configuration) for both approaches, asymptotic (left) and three-dimensional (right).
The circumferential wind is increased by $\sim3\usk\meter\per\second$ and $\sim1.5\usk\meter\per\second$ due to the anti-parallel alignment of the diabatic heat release for the asymptotic and three-dimensional simulation, respectively.}
\label{fig:MWS_stag_intens}
\end{figure}

In contrast to figure~\ref{fig:MWS_stag}, where we did not see any sizeable impact of diabatic heating on the mean circumferential velocity, figure~\ref{fig:MWS_stag_intens} displays a clear increase for both the asymptotic and three-dimensional equations.
We notice that the intensification is more effective in the asymptotic simulation ($\sim$ 3 m/s) compared to the three-dimensional simulation ($\sim$ 1.5 m/s), even though the heating amplitude of the asymptotic simulation is tuned to meet the three-dimensional simulation.
This may be caused by the fact that the centerline in the three-dimensional simulations is a reconstruction from the flow field and hence affected by inaccuracies leading to non-optimal alignment of the heating dipole.
Furthermore, even with rather high resolution the three-dimensional simulation is affected by numerical damping leading to a decrease of the centerline tilt over time which further reduces the impact (cf. eqs. \eqref{eq:TempEvoluUthetaAsymm} and \eqref{eq:urstar0}) even with $\vect Q_{1,\theta_0}$ computed by \eqref{eq:Q1_theta_0} with values of $\partial_z \vect X$ just before the heating.
Figure~\ref{fig:compareTilt} reveals the tilt dynamics without and with diabatic heating for both simulation approaches. In the three-dimensional simulations the tilt is more distorted than in the asymptotic simulations and the effect on the centerline tilt is less pronounced.
However, the overall behavior is comparable between the two simulations.

\begin{figure}[htbp]
 \centering
 \includegraphics[width=0.49\textwidth]{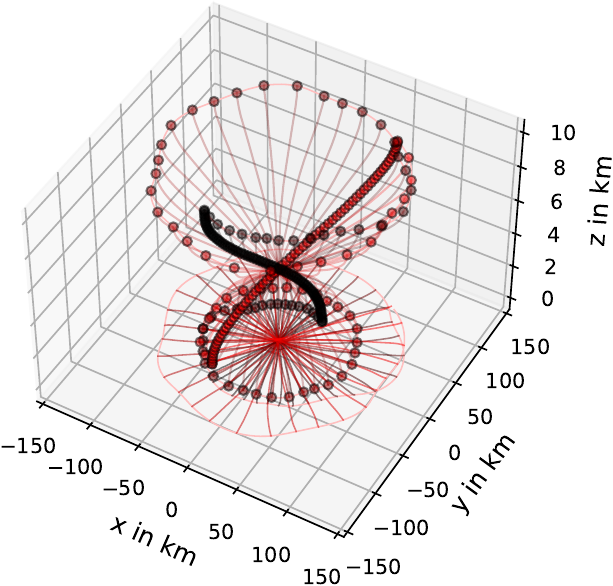}
 \hfill
 \includegraphics[width=0.49\textwidth]{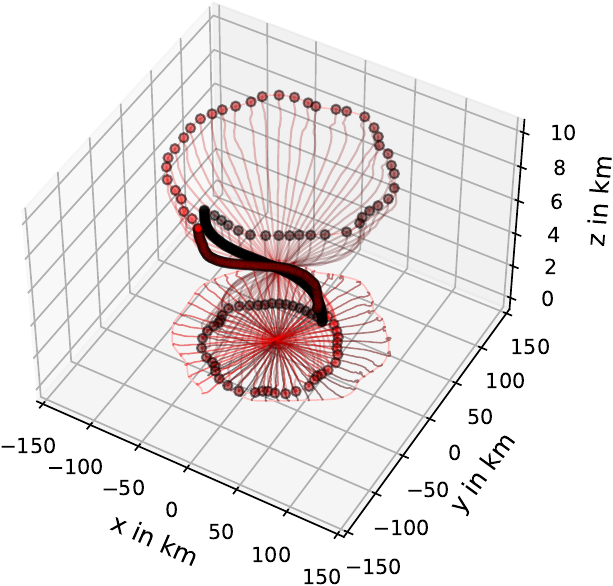}
 \caption{Same as figure \ref{fig:CenterlineMotionStagnation} with $\theta_0 = \pi/2$ (intensification configuration).}
\label{fig:CL_intens}
\end{figure}

From the discussion in subsection \ref{ssec:HeuristicsCenterlineMotion} we concluded that a heating pattern with $\theta_0 = \pi$ would lead to an increase of the centerline tilt.
This behavior could be validated in the current situation as seen in figure~\ref{fig:CL_intens} for both simulations, asymptotic and three-dimensional.
As the increase in tangential velocity is not as efficient for the three-dimensional simulation as it is for the asymptotic one, after the period of heating, the centerline precesses with higher angular frequency in the asymptotic case.
Again, this may be due to inaccuracies in the three-dimensional simulation when computing and aligning the heating dipole.

\begin{figure}[htbp]
\centering
\includegraphics[height=0.37\textwidth]{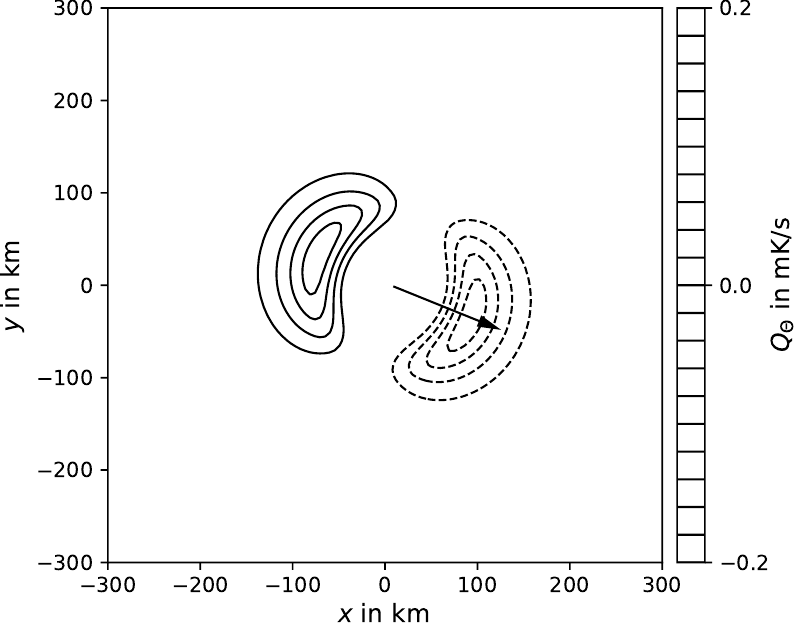}
\hfill
\includegraphics[height=0.37\textwidth]{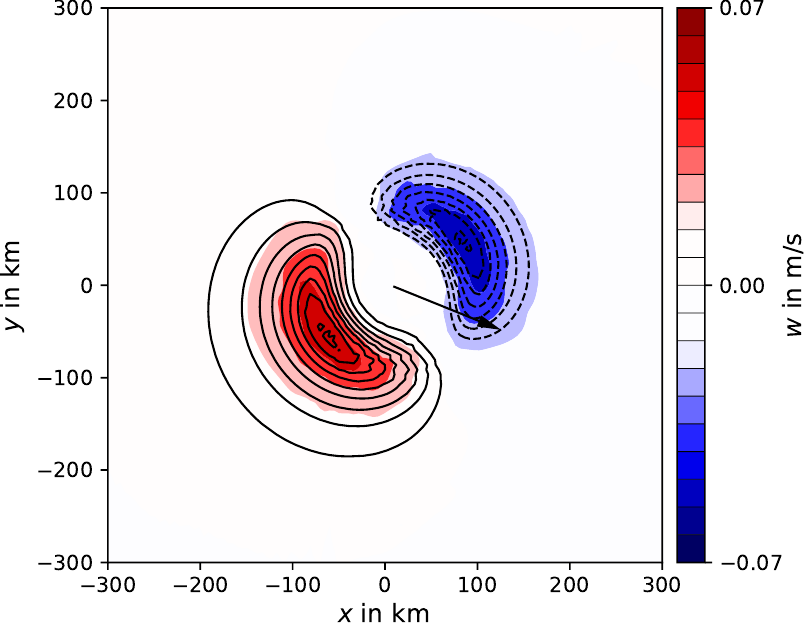}
\caption{Same as figure \ref{fig:w_stag} with $\theta=\pi$ (intensification configuration).}
\label{fig:w_stag_intens}
\end{figure}

In section \ref{ssec:SpinUpMechanism} we discussed the intensification mechanism as result of the effective vertical velocity dipole resulting from the superposition of both adiabatic and diabatic contributions.
In the current situation the diabatic vertical velocity dipole is oriented by $\theta_0=\pi$ while the adiabatic dipole is oriented by $-\pi/2$.
As both have comparably the same amplitude we would expect $-3\pi/4$ rotation between tilt and resulting vertical velocity which is verified by results shown in the right panel of figure~\ref{fig:w_stag_intens}.

\begin{figure}[htbp]
\centering
\includegraphics[height=0.45\textwidth]{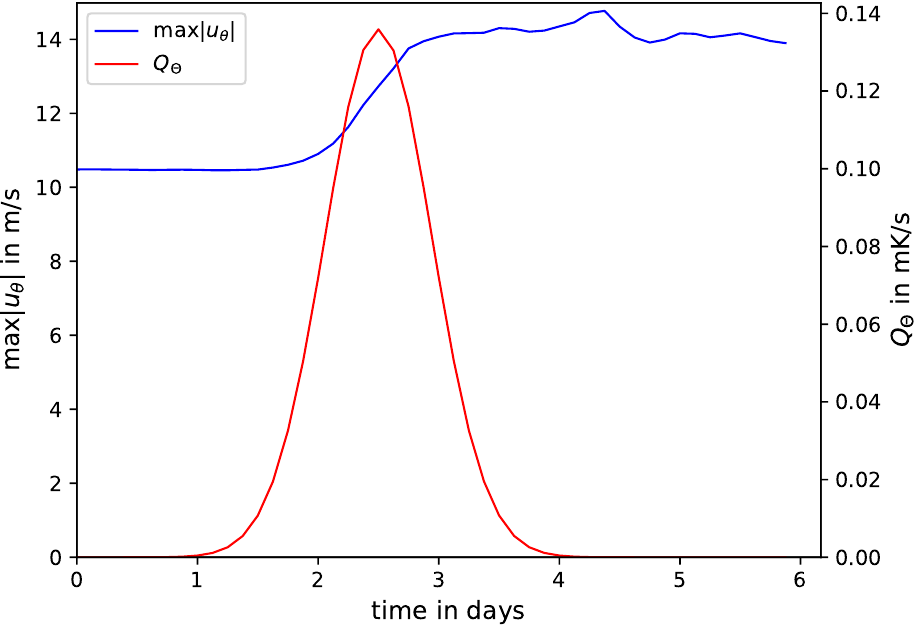}
\caption{Same as figure \ref{fig:MWS_stag_intens}, right panel, for the three-dimensional intensification experiment with $\theta_0 = \pi$ and an extended heating phase between day 1 and 4.
}
\label{fig:MWS_intens}
\end{figure}

Although the intensification is rather weak, there is evidence that the orientation of the asymmetry of the diabatic heat release matters for the evolution of the circumferential velocity.
In fact, by allowing for stronger heating by increasing the duration of the heat pulse the effect on the tangential velocity is stronger (see figure~\ref{fig:MWS_intens}), but it also affects the structure of the centerline in a more profound manner.

\begin{figure}
 \centering
 \includegraphics[height=0.45\textwidth]{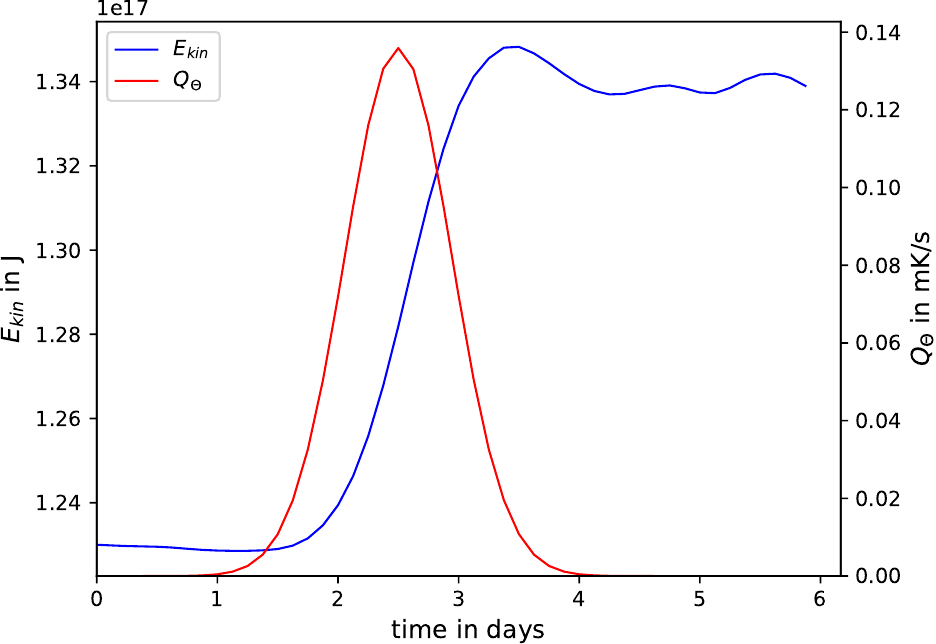}
 \caption{Same as figure~\ref{fig:MWS_intens} with the blue line representing the time series of the domain-integrated kinetic energy.
 }
 \label{fig:Energy_intens}
\end{figure}

Figure~\ref{fig:Energy_intens} also demonstrates that the system's kinetic energy increases as a consequence of the imposed asymmetric heating in line with our findings of section~\ref{ssec:EnergyBudget}.

\subsection{Attenuation}
\label{ssec:attenuation}

The final experiment of this work consists in switching the heating dipole pattern to a configuration where we expect attenuation of the vortex and vertical alignment of the centerline.
Following equation \eqref{eq:urstar0} attenuation corresponds to $\theta_0=0$, \ie, positive heating in the direction of the centerline tilt.
Again, tilt amplitude and tangential velocity of eq. \eqref{eq:Q1_theta_0} are set to initial values to avoid non-linear feedback and restricted by an amplitude factor to act over a short interval only.

\begin{figure}[htbp]
 \centering
 \includegraphics[width=0.49\textwidth]{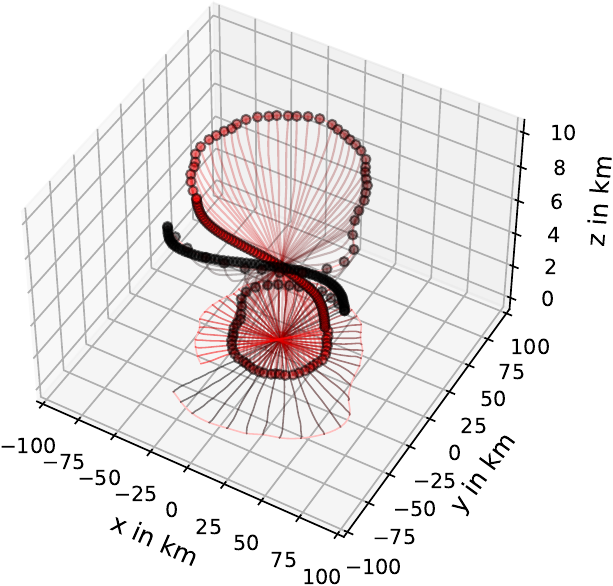}
 \hfill
 \includegraphics[width=0.49\textwidth]{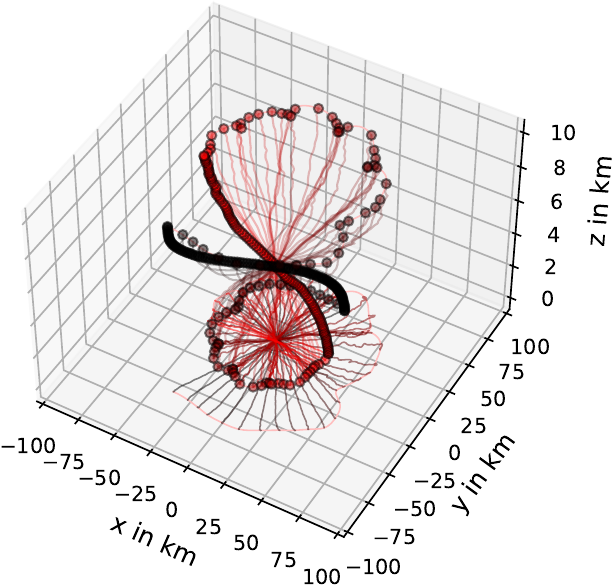}
 \caption{Same as figure \ref{fig:CenterlineMotionStagnation} with $\theta=0$.}
 \label{fig:CL_atten}
\end{figure}

\begin{figure}
 \centering
 \includegraphics[height=0.45\textwidth]{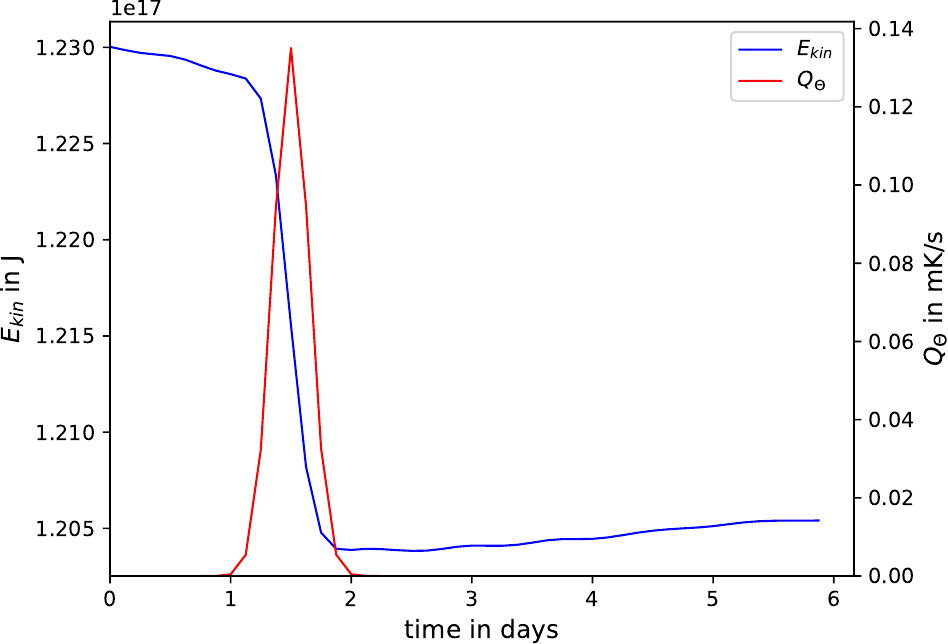}
 \caption{Same as figure \ref{fig:Energy_intens} with $\theta_0 =0$ (attenuation configuration).}
 \label{fig:Energy_atten}
\end{figure}

In figure~\ref{fig:CL_atten} for both, asymptotic and three-dimensional simulations, the centerline aligns when forced by heating.
In addition, figure~\ref{fig:Energy_atten} demonstrates the reduction of integrated kinetic energy due to the attenuating heating dipole.

\subsection{Summary of tilt dynamics}

\begin{figure}[htbp]
  \centering
  \includegraphics[width=0.9\textwidth]{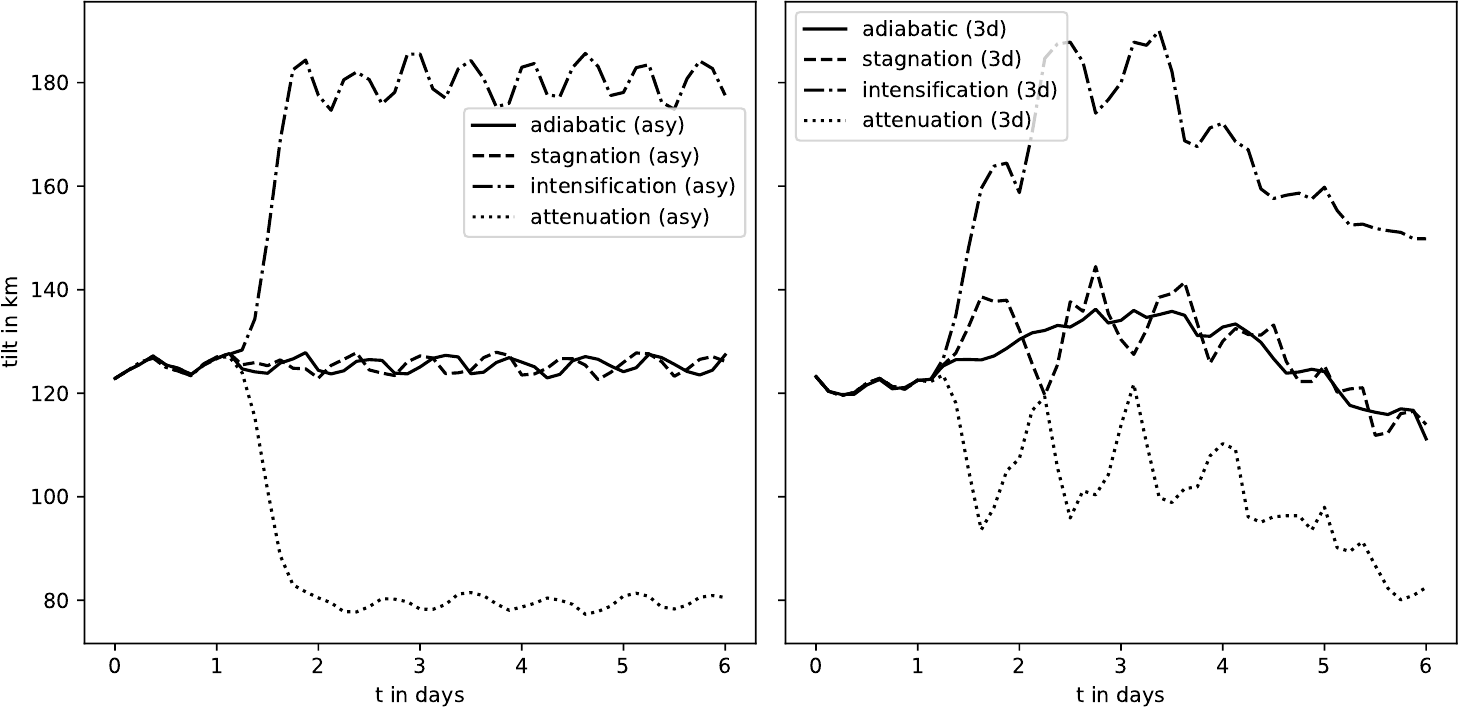}
  \caption{Time series of tilt amplitudes for the different experiments.
  Both panels range over the same time period but the three-dimensional simulations are preceeded by 4 days of initial balancing.}
  \label{fig:compareTilt}
\end{figure}

We want to emphasize the effect of asymmetric diabatic heating on the centerline tilt by analyzing the time series of tilt amplitude as measured by an $L_2$-norm:
\begin{equation}
  \left\lVert \pderiv{\vect X}{z} \right\rVert := \int\limits_0^{z_{top}} \left( \sqrt{\pderiv{\vect X}{z}^2} \right) \, dz, \qquad z_{top} = 10 \usk\kilo\meter,
\end{equation}
With figure~\ref{fig:compareTilt} this quantity is plotted for all types of experiments performed in the course of this work.
It confirms on the one hand, that the orientation of the diabatic heating dipole correlates with changes in the tilt amplitude.
Aligning the heating dipole with the tilt (attenuation case) leads to decreasing the tilt, \ie, the vortex aligns.
This situation turns around if the heating dipole is rotated 180$\degree$ (intensification case) in that the tilt further increases.
The stagnation configuration (tilt and heating dipole are perpendicular) leads to no significant alterations from the adiabatic behavior for the asymptotic simulations.

On the other hand, we see that the three-dimensional representation of the initial data does not project as well onto the first eigenmode of the centerline as in the asymptotic cases.
Besides features which oscillate on a time scale of less than one day all three-dimensional simulations exhibit an additional oscillation on the scale of roughly six days.
However, in every case, asymptotic and three-dimensional, the effects of activating an asymmetric diabatic heating dipole are superimposed onto the adiabatic reference.
The stagnation simulation follows the adiabatic reference on both cases but exhibits slight distortions in the three-dimensional case.
As discussed before this is the reason for restricting the diabatic heating to a short time period as this effect would increase for longer periods.
However, for both experiments, intensification and attenuation, the tilt amplitude follows approximately the adiabatic reference curve on an increased and lowered level, respectively.

We argue that the response of asymmetric diabatic heating for the three-dimensional simulations is not as direct as it is for the asymptotic analog simulations.
This is probably due to imbalances excited through the diabatic heating, \eg, by misalignment of the heating dipole, and limits the efficacy of the heating in \eqref{eq:Q1_theta_0} in influencing tilt and circumferential velocity.
Nonetheless, in all cases the intensification configuration increases the centerline tilt while the attenuation configuration decreases it.

The analysis above corroborates our initial statement that the orientation of a purely dipolar diabatic heating pattern intensifies and shears a atmospheric vortex apart in the anti-parallel orientation of heating dipole and tilt while it attenuates and aligns the vortex vertically in the parallel orientation.



\section{Conclusions and outlook}
\label{sec:Conclusions}

With the present work we have extended the results of \citet{PaeschkeEtAl2012} to vortex Rossby numbers larger than unity corresponding to wind speeds of $\mathcal O(30\usk\meter\per\second)$  and showed that the principal structure of the reduced asymptotic model equations is retained in this limit. We found that the validity of the equations holds from the Gradient-wind regime up to (modest) cyclostrophic vortices. This corresponds to hurricanes of strength H1 on the Saffir-Simpson scale.

Current models of tropical cyclone intensification rely on organized symmetric heating in an upright vortex \citep[][and references therein]{SmithMontgomery2017}.
Observations show that in incipient tropical storms the level of organization of convection is weak compared to that of mature storms.
Our findings of leading-order effects of asymmetric diabatic heating both on the strength of the primary circulation and on the vortex tilt may add a new route of acceleration from tropical storms to mature hurricanes.

Here, we have focused on highlighting the potential effect of asymmetric diabatic heating release on the mean tangential velocity, which we argued, can be of the same order of magnitude as symmetric diabatic heat release.
Although the gain of horizontal wind was limited as we restricted to heating configurations which maintained the overall flow structure which the asymptotic analysis was based on, we expect to see potentially stronger efficiencies in nature due to the self-regulation of moist convection in a sheared environment.
Moist thermodynamics has been replaced here, however, by artificial diabatic source and sink terms neglecting effects of water phase transitions. As we argued, the resulting asymmetric pattern of vertical velocity is the driver for intensification/attenuation.
In ongoing research, to be published elsewhere in the future, we find that mean vertical mass fluxes of an ensemble of convective towers can provide for similar effects.
Thus, despite its somewhat artificial setup, the present study does reveal an interesting physical mechanism.

In this work we did not discuss the interaction of environmental shear with the TC. Observations of TCs just before rapid intensification often show a phase of relatively stationary configuration of background wind shear \citep{RyglickiEtAl2018a}, vortex tilt, and asymmetries of convection \citep{SmithZhangMontgomery2017}.
Further, we explicitly restricted to initially barotropic velocity distribution to avoid interaction due to baroclinicity.
Still, these interactions may find an explanation in the framework of the present asymptotic model and are subject to current investigation.




\begin{acknowledgement}
T.D., A.P., and R.K.'s work has been funded by Deutsche Forschungsgemeinschaft (DFG) through grant CRC~1114 ``Scaling Cascades in Complex Systems'', Project Number 235221301, Project (C06) ``Multi-scale structure of atmospheric vortices'', and by the Helmholtz Society of Research Institutions for funding through the ``GeoSim'' Graduate College.
The authors also thank the European Centre for Medium Range Weather Forecast for supporting this work by R.K's ECMWF research fellowship as well as hosting T.D., A.P., and R.K. for research stays.
The authors acknowledge the North-German Supercomputing Alliance (HLRN) as well as the German Climate Computing Center (DKRZ) for providing HPC resources that have contributed to the research results reported in this paper.
The authors also gratefully thank Olivier Pauluis, Remi Taullieux, and Mike Montgomery for many fruitful discussions which have helped strengthen our interpretation of the asymptotic results.
T.D. and R.K. further thank Sundararaman Gopalakrishnan, Frank Marks, Paul Reasor, and Dave Nolan for their hospitality and insightful discussions during a research stay at NOAA/HRD and University of Miami.
NCAR is sponsored by the National Science Foundation.
\end{acknowledgement}


\appendix
\section{Governing equations in the co-moving coordinates}
\label{sec:TransformedEquations}

Transforming (\ref{eq:CompressibleFlowEquationsDimless}) to the vortex-centered
coordinates from section \ref{ssec:TiltCoordinates}
using \eq{eq:DerivativeTransformations} and defining
$\VC \equiv \sqeps\,\pptext{\XC}{t}$ and $\vrel = u_r\, \verad + u_{\theta}\, \vetheta$
we find
\begin{subequations}
  \label{Co-movingNondimensionalCompressibleFlowEquations}
  \begin{align}
    \begin{split}
    \label{Co-movingHorizontalMomentumEquation}
    \pp{(\VC+\vrel)}{\tv}
    + \frac{1}{\sqeps} \vrel \cdot\gradhorcyl{\vrel}
    + \frac{\w}{\sqeps^3} \left[\pp{}{z}
              - \tilt\cdot\gradhorcyl
         \right] (\VC+\vrel)
    &\\
    +\ \frac{1}{\sqeps^7} \frac{1}{\rho} \, \gradhorcyl{p}
    + \frac{1}{\sqeps^2} f\, \vk \times (\VC+\vrel)& =
    0 \, ,
    \end{split}
    \\
    \begin{split}
    \label{Co-movingVerticalMomentumEquation}
    \pp{w}{\tv}
    + \frac{1}{\sqeps} \vrel \cdot \gradhorcyl\statt{\vp}{w}
    + \frac{\w}{\sqeps^3} \left[\pp{}{z}
            - \tilt \cdot \gradhorcyl
      \right] \w
    &\\
    +\ \frac{1}{\sqeps^{10}}\,
    \left(\frac{1}{\rho}\,
          \left[\pderiv{}{z}
              - \tilt\cdot\gradhorcyl
          \right]\p
    + 1
    \right)
    \,
    &
    = 0\, ,
    \end{split}
    \\
    \begin{split}
    \label{Co-movingContinuityEquation}
    \pp{\rho}{\tv}
    + \frac{1}{\sqeps} \gradhorcyl\cdot \left(\rho \vrel\right)
    + \frac{1}{\sqeps^3}\left[\pderiv{}{z}
        - \tilt\cdot\gradhorcyl
      \right] (\rho\w)
    & =
    0\, ,
    \end{split}
    \\
    \label{Co-movingTransportEquationpontetialTemperature}
    \pp{\Theta}{\tv}
    + \frac{1}{\sqeps} \vrel \cdot\gradhorcyl{\Theta}
    + \frac{\w}{\sqeps^3} \left[\pderiv{}{z}
               - \tilt\cdot\gradhorcyl
          \right] \Theta
    &=
    Q_{\Theta}\, .
  \end{align}
\end{subequations}
%


\section{Full second order horizontal momentum balances}
\label{sec:FullSecondOrderMomentum}

\begin{subequations}\label{eq:HormomSecondApp}
\begin{eqnarray} \label{mom_r_32}
\frac{\uthet\zero}{\rmeso}\pp{\urad\order{2}}{\theta}
- \frac{2 \uthet\zero\uthet\order{2}}{\rmeso}
- \frac{(\uthet\order{1})^2}{\rmeso}
+ \ \vetheta \cdot \pderiv{\XC\order{0}}{z} \frac{\wc\order{0}\uthet\zero}{\rmeso}
\hspace*{0.7cm}
  \\  \nonumber
+\ \frac{1}{\rho_{0}}\pp{\pc\order{6}}{\rmeso}
-  \frac{\rho_{1}}{\rho_0^2}\pp{\pc\order{4}}{\rmeso}
- f_0\, \uthet\order{1}
= 0
\end{eqnarray}
\begin{eqnarray} \label{mom_theta_32}
\pp{\uthet\zero}{t}
+ \wc\order{0} \pp{\uthet\zero}{z}
+ \urad\order{2}\left( \pp{\uthet\zero}{\rmeso} + \frac{\uthet\zero}{\rmeso} \right)
+ \frac{\uthet\zero}{\rmeso} \pp{\uthet\order{2}}{\theta}
\hspace*{1.0cm}
\\ \nonumber
-\ \wc\order{0} \verad \cdot\pderiv{\XC\order{0}}{z}\pp{\uthet\zero}{{\rmeso}}
+  \frac{1}{\rho_0 \rmeso}  \pp{\pc\order{6}}{\theta}
= 0 \hspace*{0.4cm}
\end{eqnarray}
\end{subequations}
%


\section{Derivation of the kinetic energy budget \eq{eq:EKinBalance}}
\label{sec:KineticEnergyBudgetApp}

We start from \eq{eq:MomThetaTwoAverageAsymmetriesSplitOff}, which is equivalent to
(4.21) in \citep{PaeschkeEtAl2012} for $f_0 = 0$. This is verified straightforwardly
by using $u_{r,0}\order{2} = u_{r,00}\order{2} + u_{r,*}\order{2}$. The equation is
multiplied by $\rmeso\rho_0 \uthet\order{0}$, and we use the mass conservation law
in the form of eq.\ \eq{anelastic}, \ie,
\begin{equation}\label{eq:MassConservationApp}
\left(\rmeso \rho_0 u_{r,00}\order{2}\right)_{\rmeso}
+ \left(\rmeso \rho_0 w_0\order{0}\right)_z = 0\,,
\end{equation}
to generate the advective transport terms of kinetic energy in conservation form.
We let
\begin{equation}
\ekin = \rho_0 {\uthet\order{0}}^2 / 2
\end{equation}
and obtain
\begin{equation}\label{eq:EKinBalanceAppI}
\left(\rmeso e_k\right)_t
+ \left(\rmeso u_{r,00}\order{2} \ekin\right)_{\rmeso}
+ \left(\rmeso w_0\order{0} \ekin\right)_z
= - \rmeso \left(u_{r,00}\order{2} + u_{r,*}\right) \pp{p\order{4}}{\rmeso}\,.
\end{equation}
Focusing on the right-hand side of this equation, we rewrite the first term as
\begin{equation}\label{eq:PressureWorkTerms}
\begin{array}{rcl}
\dss \rmeso u_{r,00}\order{2}\pp{p\order{4}}{\rmeso}
  & =
    & \dss \left(\rmeso u_{r,00}\order{2}p\order{4}\right)_{\rmeso}
      + \left(\rmeso w_{0}\order{0}p\order{4}\right)_{z}
      - \rmeso \rho_0 w_{0}\order{0} \left(\frac{p\order{4}}{\rho_0}\right)_z
      \\
  &
    & \dss - \frac{p\order{4}}{\rho_0}
      \left[\left(\rmeso \rho_0 u_{r,00}\order{2}\right)_{\rmeso}
          + \left(\rmeso \rho_0 w_{0}\order{0}\right)_{\rmeso}
      \right]\,.
\end{array}
\end{equation}
The square bracket vanishes according to \eq{eq:MassConservationApp}, while we observe
that by combining the axisymmetric part of the hydrostatic balance in
\eq{vertmom2} with the equation of state in \eq{state42} to replace
$(p\order{4}/\rho_0)_z$, and \eq{w52} to replace $w_0\order{0}$ one finds
\begin{equation}\label{eq:SymmetricHeatingPart}
\rmeso\rho_0 w_0\order{0}\left(\frac{p\order{4}}{\rho_0}\right)_z
= \rmeso \rho_0 \frac{Q_{\Theta,0}\order{0}}{d\Theta_2/dz} \frac{\Thetac\order{4}_0}{\Theta_0}\,.
\end{equation}
Next we rewrite the second term on the right of \eq{eq:EKinBalanceAppI}
using the definition of $u_{r,*}\order{2}$ in \eq{eq:UrStar}, and the
first Fourier modes of the vertical momentum balance in \eq{vert11newtest2}
to find
\begin{equation}\label{eq:AsymmetricHeatingPart}
\rmeso u_{r,*}\order{2} \pp{p\order{4}}{\rmeso}
= {\frac{1}{2}} \rmeso\rho_0\ \vect{w}_1\cdot \frac{\vect{\Theta}_1\order{4}}{\Theta_0}
= {\frac{1}{2}} \rmeso\rho_0\ \frac{\vect{Q}_{\Theta,1}\order{0}}{d\Theta_2/dz}
  \cdot \frac{\vect{\Theta}_1\order{4}}{\Theta_0}\,.
\end{equation}
To obtain the second equality we have used the asymmetric WTG-law from
\eq{w5211} and the fact that the second term in that equation contributes
a component to $\vect{w}_1$ that is orthogonal $\partial\vect{X}\order{0}/\partial z$,
and thus also orthogonal to $\vect{\Theta}_1\order{4}$.

Insertion of \eq{eq:PressureWorkTerms}--\eq{eq:AsymmetricHeatingPart} generates
the desired equation \eq{eq:EKinBalance}.


\section{Relation to Lorenz' theory of Available Potential Energy}
\label{sec:APE}

We want to strengthen the fact that our outlined theory is compliant with Lorenz' concept of available potential energy (APE) \citep{Lorenz1955}.
We will see, under certain assumptions, that the kinetic energy generation of equation \eqref{eq:EKinBalance} is identical to Lorenz' expressions of APE generation and conversion to kinetic energy.

Therefore, we start with his equations (16), (20), (17), (18):
\begin{align}
 \dd{\mean A}{t} &= -C + G \,, \\
 \dd{\mean K}{t} &= C      \,, \\
 C              &= -\frac{R}{g} \int\limits_0^\psurf \frac{1}{p} \mean{T\omega}\, dp \label{eq:ConversionAPE}\\
 G              &=  \frac{1}{g} \int\limits_0^\psurf \frac{\Gamma_d}{\Gamma_d - \mean\Gamma} \frac{\mean{T'Q'}}{\mean T} \, dp \,.\label{eq:GenerationAPE}
\end{align}
Here, $\mean A$ is the average available potential energy, $\mean K$ the mean kinetic energy, $C$ the conversion of APE to kinetic energy, $G$ the generation of APE (all per unit surface), $\psurf$ the surface pressure $\omega = \dd{p}{t} = \dd{p}{z} w = -g \mean\rho w$ the vertical velocity in pressure coordinates, $\mean\Gamma= \pp{\mean T}{z}$ the  lapse rate of temperature, $\Gamma_d=g/\cp$ the dry-adiabatic lapse rate, $Q = \cp \Theta Q_\Theta$ the heat rate, $\mean{(\cdot)}$ the horizontal mean, and $\devi{(\cdot)}$ the deviation from the mean.
In our case we neglect friction, hence equation (21) of \citet{Lorenz1955} is trivially zero.

We can make use of the results from above, as well as  those from \citet[][and references therein]{PaeschkeEtAl2012}, that the leading-order expansion modes of the thermodynamical quantities are horizontally homogeneous. For example, for potential temperature we have
\begin{align}
 \mean\Theta &= \rfr{T}\left( \Theta_0 + \delta^2\Theta_2 + \bigoh{\delta^4} \right) \\
 \devi\Theta &= \rfr{T}\left( \delta^4\Theta\order{4} + \bigoh{\delta^5} \right)
\end{align}
Furthermore, we can find derived expression:
\begin{align}
 \mean T = \mean \Theta \mean \pi \\
 \frac{\devi T}{\mean T} = \frac{\devi \Theta}{\mean \Theta} + \frac{\devi \pi}{\mean \pi}
\end{align}
Additionally, we make use of the circumstance, that the mean of an asymmetric field is zero.

With that we get for equation \eqref{eq:GenerationAPE}
\begin{equation}
 G = g\int\limits_0^H \frac{\mean \rho}{\mean \Theta \dd{\mean \Theta}{z}} \mean{(Q_\Theta \devi{\Theta})} \, dz\,, \label{eq:GenerationConversionAPE}
\end{equation}
which is the the right-hand side of equation \eqref{eq:EKinBalance} in physical dimensions, integrated over the whole domain, where $Q_\Theta$ is decaying sufficiently fast.

With \eqref{w52} we get for the resulting vertical velocity
\begin{equation}
 w = \frac{1}{\dd{\Theta}{z}} \left( Q_\Theta - \frac{u_\theta}{r} \vect \Theta'_1 \cdot \vect e_r \right)
\end{equation}
in physical units.
With that we get for the horizontal mean $\mean{(\devi T \omega)}$
\begin{equation}
 \mean{(\devi T \omega)} = -\frac{g}{\mean R} \frac{\mean p}{\mean \Theta \dd{\mean\Theta}{z}} \mean{\left( \devi \Theta Q_\Theta \right)}\,.
\end{equation}
Inserted into \eqref{eq:ConversionAPE} we get for $C$ the same expression as for $G$ in \eqref{eq:GenerationConversionAPE}.

Finally, we conclude, that the generation of kinetic energy is equivalent to the generation of APE in Lorenz' theory and furthermore, that the expressions of generation APE and conversion into kinetic energy are identical for the setup of an asymmetrically heated vortex.
All the APE, which is generated by heating is converted immediately into kinetic energy.


\section{The centerline equation of motion}
\label{sec:CenterlineMotion}

As preparation for appendix \ref{sec:NumericsAsymptotics} give more details on the centerline equation of motion \eqref{eq:TempevolutionCenterlineIntro} and provide all the necessary information to close the system of equations (in conjunction with \eqref{eq:TempevoluthetaIntro}).

We will present the missing terms necessary to close equation \eqref{eq:TempevolutionCenterlineIntro} in section \ref{ssec:AsymPaeschke}, provide a split into adiabatic and diabatic contributions in section \ref{ssec:AsymSplit} and further deepen the analysis of this equation in section \ref{ssec:CharCenterline} for being able to construct a stable and efficient numerical scheme to solve the asymptotic equations.

\subsection{Formulation of Paeschke et al. (2012)}
\label{ssec:AsymPaeschke}

In the course of this work we argued that the equations derived by \citet{PaeschkeEtAl2012} are valid in the vanishing-Coriolis case, \ie, $f_0 \rightarrow 0$.
The structure of the equations stays essentially the same, only terms proportional to $f_0$ are to be dropped.

For completeness below we present the remaining expressions for $\vect M_1$ and $\vPsi$ necessary to solve the system \eqref{eq:TempevoluthetaIntro} and \eqref{eq:TempevolutionCenterlineIntro}:

\begin{align}
  \vect M_1 = \frac{f^2}{4\pi\rho_0 \Gamma} \partial_z
    \left( \frac{\rho_0 \Gamma^2}{\Theta_2'} \partial_z \vect X \right)\,
\end{align}
$\vPsi = L[\Kcal] = \Rotmat_{\pi/2} L[\Kcal_1]$ depends the following expressions:

\begin{subequations}
  \begin{align}
    \Kcal_1   &= \Hcal_1 + \Itcal_1 + \Jcal_1 + \Qcal_1\,, \label{eq:Kcal}\\
    \Hcal_1   &= \partial_r \left( r \vect w_1 \partial_z u_\theta \right)\,, \\
    \Itcal_1  &= \Ical_1 + H_s(r-1) \frac{1}{r^2} \vect I_1\,, \\
    \Ical_1   &= r\left( \zeta + f \right) \vect W_1\,, \\
    \vect I_1 &= \frac{\Gamma}{2\pi} \Rotmat_{-\pi/2} \vect M_1\,, \\
    \Jcal_1   &= (\partial_r \vect \phi_1) (r\partial_r \zeta)\,, \\
    \Qcal_1   &= (w_0 \frac{u}{r} - \partial_r(r w_0 \partial_r u_\theta)) \partial_z \vect X\,, \label{eq:Qcal}
  \end{align}
  \label{eq:Psi_components}
\end{subequations}
where $L[\cdot]$ is an integral operator,
\begin{equation}
  L[\Kcal] = \frac{\pi}{\Gamma} \int\limits_0^\infty r \Kcal(r) \,dr\,, \\
\end{equation}
and $\Rotmat_{\theta_0}$ denotes the matrix of two-dimensional rotation by an angle $\theta_0$.

Equations \eqref{eq:Kcal} -- \eqref{eq:Qcal} involve terms which are resolved in terms of $u_\theta$, $\vect X$, $Q_0$ and $\vect Q_1$:
\begin{align}
  \vect W_1    &= -\frac{1}{\rho_0} \partial_z (\rho_0 \vect w_1) \\
  w_0          &= \frac{Q_0}{\Theta'_2} \\
  \vect w_1    &= \frac{1}{\Theta_2'} \left( \vect Q_1 + W \Rotmat_{-\pi/2} \partial_z \vect X \right) \label{eq:Asymw1}\\
  W            &= \frac{u_\theta}{r} \left( \frac{u_\theta^2}{r} + fu_\theta \right) \\
  \zeta        &= \frac{1}{r}\pderiv{(r u_\theta)}{r} \\
  \vect \phi_1 &= -r \int\limits_r^\infty \frac{1}{\bar r^3}
    \int\limits_0^{\bar r} \bar{\bar r}^2 \vect R_1 \, d\bar{\bar r} \, d\bar r \\
  \vect R_1    &= \vect W_1 + \frac{1}{2}(\partial_r w_0) \partial_z \vect X
\end{align}
The expressions above give rise to the reformulated equation of centerline motion:
\begin{align}
  \partial_t \vect X = \vect u_s +
    \frac{1}{2}\ln (\delta) \vect k \times \vect M_1 - L[\Kcal_1]
\end{align}

\subsection{Split into diabatic and adiabatic contributions}
\label{ssec:AsymSplit}

To reveal more clearly the structure of the centerline evolution equation,
we recall the formula \eq{w5211} for the vertical velocity Fourier modes,
which separates diabatic from adiabatic effects. Rewriting this formula
in the dipole vector notation we have
\begin{align}
  \vect w_1        &= \vect w_{1,\dia} +\vect w_{1,\ad}\,, \\
  \vect w_{1,\dia} &= \frac{1}{\Theta_2'} \vect Q_1\,, \\
  \vect w_{1,\ad}  &= \frac{1}{\Theta_2'} W \Rotmat_{-\pi/2} \partial_z \vect X
    =: \Rotmat_{-\pi/2} \hat w \vect X\,,
\end{align}
and realize that the adiabatic part is a linear (differential) operation on $\vect X$:
\begin{equation}
  \hat w = \frac{W}{\Theta_1'}\partial_z
\end{equation}
From the equations in the previous subsection we see that by linearity of the expressions in $\vect w_1$ we can assemble $\Kcal$ (and ultimately $\vPsi$) by linear superpositions of linear operations on $\vect X$ (operators are symbolically denoted by $\hat{~}$) and in general nonlinear diabatic expressions.

$\Hcal_1$ then becomes:
\begin{align}
  \Hcal_1
    &= \Hcal_{1,\dia} + \Rotmat_{-\pi/2} \hat{\mathcal H} \vect X \\
  \Hcal_{1,\dia}
    &= \partial_r \left( r \vect w_{1,\dia} \right) \\
  \hat{\mathcal H} \vect X
    &= \partial_r \left( (\hat w \vect X) \partial_z u_\theta \right)
\end{align}
We first need to evaluate the expression for $\vect W_1$,

\begin{align}
  \vect W_1
    &= \vect W_{1,\dia} + \Rotmat_{-\pi/2} \hat W \vect X\\
    \vect W_{1,\dia}
    &= -\frac{1}{\rho_0} \partial_z (\rho_0 \vect w_{1,\dia}) \\
  \vect W_{1,\ad}
    &= -\Rotmat_{-\pi/2} \frac{1}{\rho_0} \partial_z (\rho_0 \hat w \vect X) := \Rotmat_{-\pi/2} \hat W_{1,\ad} \vect X
\end{align}
to split $\Ical_1$ accordingly:
\begin{align}
  \Ical_1 &= \Ical_{1,\dia} + \Rotmat_{-\pi/2} \hat {\mathcal I} \vect X \\
  \Ical_{1,\dia} &= r\left( \zeta + f \right) \vect W_{1,\dia} \\
  \hat {\mathcal I} \vect X &= r\left( \zeta + f \right) \hat W_{1,\ad} \vect X
\end{align}
Together with $\vect M_1$
\begin{align}
  \vect M_1 &= \frac{f^2}{4\pi\rho_0 \Gamma} \partial_z
    \left( \frac{\rho_0 \Gamma^2}{\Theta_1'} \partial_z \vect X \right) \\
            &=:\hat M \vect X
\end{align}
we get the following split for $\Itcal_1$:
\begin{align}
  \Itcal_1 &= \Ical_{1,\dia} +
    \Rotmat_{-\pi/2} \hat {\mathcal I} \vect X +
      H_s(r-1) \frac{1}{r^2} \frac{\Gamma}{2\pi} \Rotmat_{-\pi/2} \hat M \vect X \\
      &=: \Ical_{1,\dia} + \Rotmat_{-\pi/2} \hat{\tilde{\mathcal I}} \vect X
\end{align}
Performing the split of $\vect R_1$,

\begin{align}
  \vect R_1 &=
    \vect W_{1,\dia} + \Rotmat_{-\pi/2} \hat W \vect X + R_{Q,0} \partial_z \vect X \,, \\
  R_{Q,0}   &= \frac{1}{2}(\partial_r w_0)
\end{align}
$\vect \phi_1$ divides into
\begin{align}
  \vect \phi_1 &=
    \vect\phi_{1,\dia} + \Rotmat_{-\pi/2} \hat\phi \vect X + \phi_{Q,0} \partial_z\vect X \\
  \vect\phi_{1,\dia} &=
    -r \int\limits_r^\infty \frac{1}{\bar r^3} \int\limits_0^{\bar r} \bar{\bar r}^2
      \vect W_{1,\dia} \, d\bar{\bar r} \, d\bar r \\
  \hat\phi \vect X &=
    -r \int\limits_r^\infty \frac{1}{\bar r^3} \int\limits_0^{\bar r} \bar{\bar r}^2
      \hat W \vect X \, d\bar{\bar r} \, d\bar r \\
  \phi_{Q,0} &=
    -r \int\limits_r^\infty \frac{1}{\bar r^3} \int\limits_0^{\bar r} \bar{\bar r}^2
      R_{Q,0} \, d\bar{\bar r} \, d\bar r
\end{align}
and therefore $\Jcal_1$ into
\begin{align}
  \Jcal_1 &=
    \Jcal_{1,\dia} + \Rotmat_{-\pi/2}\hat{\mathcal J} \vect X + {\mathcal J}_{Q,0} \partial_z \vect X\,. \\
  \Jcal_{1,\dia} &= \partial_r (\vect\phi_{1,\dia})(r\partial_r \zeta) \\
  \hat{\mathcal J} \vect X &= \partial_r ( \hat\phi \vect X)(r\partial_r \zeta) \\
  {\mathcal J}_{Q,0} &= (\partial_r \phi_{Q,0}) (r\partial_r \zeta)
\end{align}
For $\vect Q_1$ we find the simple shorthand
\begin{align}
  \Qcal_1 &= \left( w_0 \frac{u}{r} - \partial_r(r w_0 \partial_r u) \right) \partial_z \vect X \\
          &=: {\mathcal Q}_0 \partial_z \vect X\,.
\end{align}


Our final result is identifying three different contributions to $\vPsi$, a (generally nonlinear) diabatic term, an advective term, and a linear Sturm-Liouville-type operator acting on $\vect X$:

\begin{align}
  \vPsi   &= L[\Kcal_{1,\dia}] + L[\mathcal J_{Q,0} \partial_z \vect X+  {\mathcal Q}_0 \partial_z \vect X] + R_{-\pi/2} L[\hat{\mathcal H} + \hat{\tilde{\mathcal I}} + \hat{\mathcal J}] \vect X
  \label{eq:Psi_split}
\end{align}

\subsection{Characteristic structure of the vortex centerline equation}
\label{ssec:CharCenterline}

We continue rephrasing the original centerline tendency equation to further emphasize its structure.
By trivially identifying $\mathbb R^2$ with $\mathbb C$ we symbolically transform two-dimensional vectors $\vect a = (a_x, a_y) \in \mathbb R$ to $a = a_x + i a_y \in \mathbb C$.
Operations such as ($\vect k \times \cdot)$ and $\Rotmat_{\pi/2}$ become multiplications with $i$.
By that we can identify a substructure of the equation to be of Schrödinger-type, and advective contribution and sources, generally dependent on $X$, $u_\theta$, and the coordinates $(r,z,t)$ but without any further specification:
\begin{align}
  i(\partial_t X + L[(\mathcal J_{Q,0}+  {\mathcal Q}_0) \partial_z X]) &= - \frac{1}{2} \ln \delta \hat M X - L[\hat{\mathcal H} + \hat{\tilde{\mathcal I}} + \hat{\mathcal J}] X  + \nonumber \\
    & +  i u_s - i L[\mathcal H_{Q,1} +\mathcal I_{Q,1} + \mathcal J_{Q,1}]
    \label{eq:ComplexCenterlineOperators}
\end{align}
By identifying structural components the centerline equation takes the form
\begin{align}
  i(\partial_t X + A \partial_z X) &= \hat H X + i Q + i u_s \label{eq:CLprototype}
\end{align}
Note that the left-hand sides of eq. \eqref{eq:CLprototype} takes the form of an advection operation while the right-hand side involve linear and non-linear source terms.

\citet[section 6.2]{PaeschkeEtAl2012} pointed out that the adiabatic time evolution of the vortex centerline poses an eigenproblem.
For that case the (homogeneous) centerline equation in the complex plane is written as
\begin{equation}
  i \pderiv{X_h}{t} = \hat H X_h
  \label{eq:SchroedingerEq}
\end{equation}

As the spectrum $\omega_k$ of $\hat H$ is real, \eqref{eq:SchroedingerEq} can be interpreted as a \emph{Schrödinger-type} equation, hence eigenmodes $X_k$ precess with the angular frequency $\omega_k$ in the complex plane.
For the adiabatic problem we therefore find, that
\begin{equation}
  \XC_k(t) = \Rotmat_{\omega_k t} \XC_k(t=0)\,.
\end{equation}
For the numerical experiments presented in section \ref{sec:Simulations} we use the first non-trivial eigenmode for initialization corresponding to the non-zero eigenvalue with the smallest magnitude.

\section{Numerical scheme for asymptotic equations}
\label{sec:NumericsAsymptotics}

By providing a closure $(Q_0, \vect Q_1) = F(\vect X, u_\theta, t)$ equations \eqref{eq:TempevolutionCenterlineIntro} and \eqref{eq:TempevoluthetaIntro} form a closed set of partial differential equations which in general cannot be solved analytically.
Thus, we further analyze the structure of these equations seeking for an adapted numerical method to allow for efficient and stable time integration.

\citet{Weber2011} first presented a numerical scheme for solving the coupled system \eqref{eq:TempevolutionCenterlineIntro} and \eqref{eq:TempevoluthetaIntro}.
While he followed a method-of-lines approach discretizing the spatial derivatives by fourth-order approximation to solve the resulting system of ordinary differential equations by generic integrators we try to make use of the structure revealed in appendix \ref{sec:CenterlineMotion}.
We revisited all the equations presented by \citet{PaeschkeEtAl2012} needed for closure and further performed the split of \eqref{eq:TempevolutionCenterlineIntro} into linear and non-linear contributions which lead to a quasi-Hamiltonian substructure giving rise to the numerical scheme to be presented subsection \ref{ssec:IntegrationCenterline}.
In addition to the centerline equation \eqref{eq:CLprototype} time evolution of the tangential velocity is re-written as
\begin{align}
  \partial_t u_\theta + u_{r,00} \frac{1}{r}\partial_r(ru_\theta) + w_0 \partial_z u_\theta &= - u_{r,*} \left( \frac{u_\theta}{r} + f \right) - u_{r,00}f\,\label{eq:SharpieUEvolution}
\end{align}
to identify an advection term (in polar coordinates) on the left-hand side and source terms on the right-hand side.
The integration scheme of this equation is presented in subsection \ref{ssec:IntegrationTangentialVelocity}

By our analysis we learned that the system of equations \eqref{eq:CLprototype} and \eqref{eq:SharpieUEvolution} is assembled by prototypes of partial differential equations that are i) advection equation, ii) Schrödinger equation, and iii) non-linearly coupled ordinary differential equations (source terms).

Aiming at the scope of this work we restrict to asymmetric diabatic heat release which allows to drop all terms referring to symmetric vertical and radial motions.
We further neglect background wind shear.
Hence we drop all the advective terms from eqs. \eqref{eq:CLprototype} and \eqref{eq:SharpieUEvolution} and set $u_s=0$.

\subsection{Integration of centerline}
\label{ssec:IntegrationCenterline}

For the integration of the centerline position the general solution strategy is to integrate nonlinear source terms by the trapezoidal rule and the Sturm-Liouville operator (after appropriate spatial discretization) by the implicit midpoint rule.
The choice of the latter is based on the idea of preserving unitarity during integration of the linear part of the equation.
We also dropped contributions according to shear as they are not discussed in section \ref{sec:Simulations}.
The composition of the sub-steps reads:

\begin{align}
  X^*      &= \frac{1}{2}\Delta t Q(u^n, X^n, t^n) + X^n \\
  X^{**}   &= \left(\mathbbm 1 - \frac{1}{2}i \Delta t \hat H\left(u^{n+1/2}\right)\right) X^* \\
  X^{***} &= \left(\mathbbm 1 + \frac{1}{2}i \Delta t \hat H\left(u^{n+1/2}\right)\right)^{-1} X^{**} \\
  X^{n+1}  &= \frac{1}{2} \Delta t + Q(u^{n+1}, X^{n+1}, t^{n+1}) + X^{***}
\end{align}
$\mathbbm 1$ represents the identity operator.
For the evaluation of $\hat H$ we need $u^{n+1/2}$ which we obtain by the first-order predictor $u^{n+1/2} = u^*$.

\subsection{Integration of tangential velocity}
\label{ssec:IntegrationTangentialVelocity}

The integration of the tangential velocity equation is accomplished by applying the trapezoidal rule to the source term.
Note that after dropping symmetric diabatic contributions all differential expression in equation \eqref{eq:SharpieUEvolution} vanished and we only have to integrate the non-differential source term proportional to $u_{r,*}$.
The integration scheme for one timestep $\Delta t$ reads
\begin{align}
  u^*     &= -\frac{1}{2} \Delta t u_{r,*}^n \left( \frac{u^n}{r} + f \right)  + u^n \,,\\
  u^{n+1} &= -\frac{1}{2} \Delta t u_{r,*}^{n+1} \left( \frac{u^{n+1}}{r} + f \right)  + u^{*} \,, \label{eq:uIntrfinal}
\end{align}
For clarity we dropped the ${}_\theta$ subindex.
It becomes obvious that the final step of the integration involves an implicit solution strategy as the term $u^{n+1}_{r,*}$ depends on both $X$ and $u$ at time level $n+1$.

\subsection{Coupled Integration}

The above stated integration scheme involves information from previous sub-steps for the coupled integration.
In both, explicit and implicit forcing, the equations are couples to each other.
For the implicit (finalizing) step it is necessary to iterate solving $X^{n+1}$ and $u^{n+1}$ with second order:

\begin{align*}
  X^{n+1,0}    &= X^{***} \\
  u^{n+1,0}    &= u^{**}  \\
  X^{n+1, \nu} &= \frac{1}{2} \Delta t  + Q(u^{n+1,\nu-1}, X^{n+1,\nu-1}, t^{n+1}) + X^{****} \\
  u^{n+1,\nu}  &= -\frac{1}{2} \Delta t u_{r,*}^{n+1, \nu} \left( \frac{u^{n+1,\nu}}{r} + f \right)  + u^*
\end{align*}
This integration strategy is adopted from literature on implicit methods for fluid dynamics \citep[c.f.][]{SmolarkiewiczEtAl2014, BenacchioKlein2019}.

\subsection{Details on the spatial discretization}

The equations are discretized on an equidistant grid allowing for straightforward finite-difference approximations of the derivate operators.
Boundary conditions are accommodated by a extending the grid covering the physical domain plus a ghost layer of two cells.
Solution values are stored in cell-centers while first derivate are computed typically on the corresponding faces.

Prototypical differential expressions such as $\alpha\partial_z(\beta\partial_z\psi)$ are discretized as:
\begin{equation}
  \left.\alpha\partial_z(\beta\partial_z\psi)\right|_{z=z_i} = \frac{1}{\Delta z^2} \alpha_i \left(  \beta_{i+1/2}\left( \psi_{i+1} - \psi_{i} \right)- \beta_{i-1/2}\left( \psi_i - \psi_{i-1} \right) \right)
\end{equation}
Integrals are computed via the trapezoidal rule where ghost cells are obtained by quadratic extrapolation.
We further include an option to apply hyper-viscosity to the centerline stabilizing time integration with activated diabatic heating.
Further details may be taken from the source code available on demand by the corresponding author.


\section{Details on the numerical implementation}
\label{sec:DetailsNumerical}

\subsection{Dimensional variables}

Though the derivation outlined above is carried out in terms of non-dimensional variables for the actual implementation into EULAG we used dimensional quantities of which some details will be presented in this section.
In the spirit of asymptotic analysis for reconstructing dimensional variables and formulate leading-order relations by using leading-order or next-to-leading order modes.

Before presenting specific relation which arise from the asymptotic analysis, we want to relate the expansion modes with mean background values, denoted by bars $\mean{(\cdot)}$ and perturbations, denoted by primes $\devi{(\cdot)}$.

\begin{alignat}{2}
 \mean \rho &= \rfr{\rho}\left( \rho_0 + \delta^2 \rho_2 + \delta^4 \hat \rho_4 + \bigoh{\delta^5} \right)
  \quad&\quad
   \devi \rho &= \rfr{\rho} \left( \delta^4 \hat\rho\order{4} + \bigoh{\delta^5} \right) \\
 \mean p    &= \rfr p \left( p_0 + \delta^2 p_2 + \delta^4 \hat p_4 + \bigoh{\delta^5} \right)
  \quad&\quad
   \devi p &= \rfr p \left( \delta^4 \hat p\order{4} + \bigoh{\delta^5} \right) \\
  \mean \Theta &= \rfr T \left( \Theta_0 + \delta^2 \Theta_2 + \delta^4 \hat \Theta_4 + \bigoh{\delta^5} \right)
  \quad&\quad
   \devi \Theta &= \rfr T \left( \delta^4 \hat \Theta\order{4} + \bigoh{\delta^5} \right)
\end{alignat}
Furthermore, we have
\begin{align}
 u_\theta &= \vect u \cdot \vect e_\theta = \rfr{u} \left( \frac{1}{\delta} u_\theta\order{0} + \bigoh{1} \right) \\
 w &= \hphantom{\vect u \cdot \vect e_\theta =} \, \rfr{u} \left( \delta \hat w\order{1} + \bigoh{\delta^2} \right)
\end{align}
A trivial, but useful observation is $\mean{\devi{(\cdot)}} = 0$.

\subsubsection{Pressure}

Equation \eqref{eq:HormomLeading} balances pressure gradient with radial forces.
We find the dimensional version as
\begin{equation}
 \frac{1}{\rho}\pderiv{p}{r} = \frac{u_\theta^2}{r}\,.
\end{equation}

\subsubsection{Potential temperature}

For the deviation of potential temperature from its background mean value $\mean \Theta$ we have for the first Fourier modes

\begin{equation}
 \devi{\vect \Theta}_1 = -\frac{\mean \Theta}{g} \frac{1}{\mean \rho} \pp{p}{r} \pp{\vect X}{z}\,.
\end{equation}

\subsubsection{Vertical velocity}

In general, \ie, for arbitrary heating, the vertical velocity takes the following form in physical dimensions:

\begin{equation}
 w = \frac{1}{\dd{\Theta}{z}} \left( Q_\Theta + \frac{\mean\Theta}{g} \frac{u_\theta^3}{r^2} \left( \Rotmat_{-\pi/2} \pp{\vect X}{z} \right) \cdot \vect e_r \right)
\end{equation}

\subsubsection{Diabatic heating}

Finally, we defined a heating dipole aligned with an angle $\theta_0$ relative to the tilt direction:

\begin{equation}\label{eq:Q_generalDim}
  Q_{\Theta}^{\theta_0} =
          \frac{\mean \Theta}{g} \frac{u_{\theta}^3}{r^2} \left( \Rotmat_{\theta_0} \pp{\vect X}{z} \right) \cdot \vect e_r
\end{equation}

\subsection{Convergence results}
\label{ssec:Convergence}

\begin{figure}[htbp]
\centering
\includegraphics[height=0.35\textwidth]{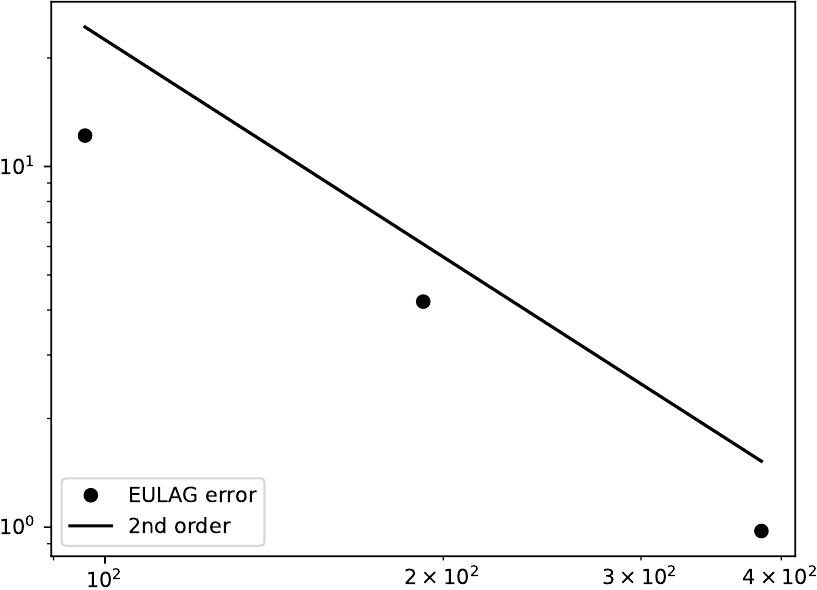}
\caption{Error convergence for EULAG simulations.
The maximum deviation of perturbation pressure from a reference solution with $768 \times 768 \times 384$ grid points at $t = 1$ day is plotted (solid line for reference).}
\label{fig:EULAGConvergence}
\end{figure}

In addition to comparing both simulation results of three-dimensional and asymptotic equations we check for convergence of each numerical schemes.
For both simulations we check for self-consistency by comparing results at increasing resultion with high-resolved reference.
Convergence of the EULAG simulations is displayed in figure~\ref{fig:EULAGConvergence} showing second-order convergence.

\begin{figure}[htbp]
 \centering
 \includegraphics[width=0.5\textwidth]{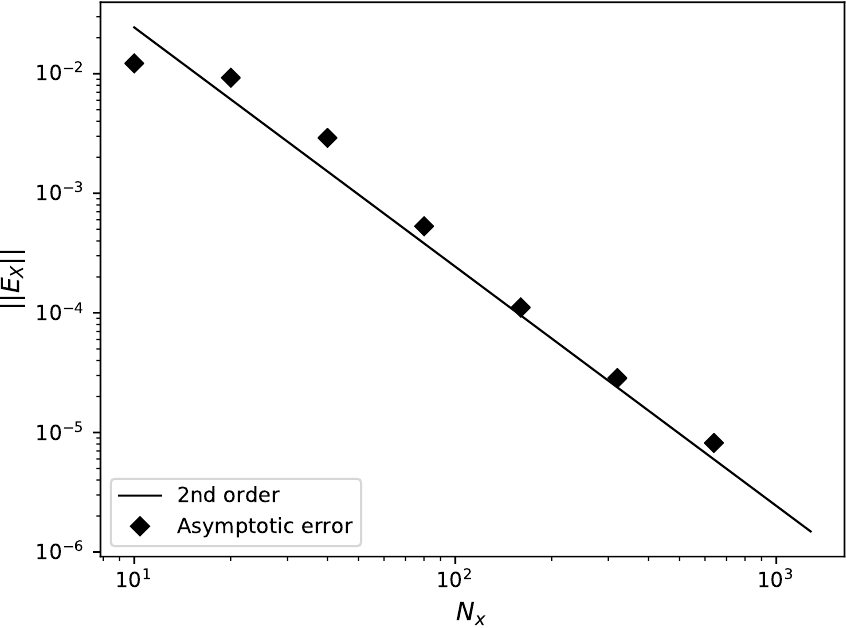}
 \caption{Convergence plot for numerical scheme solving the asymptotic equations. Error values indicated by diamond markers are computed as the difference between the current solution a reference solution with 1280 grid points.}
 \label{fig:SharpieConv}
\end{figure}

The numerical scheme solving the asymptotic equation on the other hand is tested by evolving
\begin{equation}
 \vect X(t=0) = (\cos(z\pi/z_{max}), 0)^T
\end{equation}
for $T=0.1$ (in asymptotic units) and comparing solutions of different resolutions against a reference solution with 1280 grid points.
The results are plotted in figure~\ref{fig:SharpieConv} indicating second-order convergence as expected.
%


\bigskip

\bibliographystyle{spbasic}   
\bibliography{bibliography}   

\end{document}